\documentclass[fleqn,usenatbib]{mnras}

\usepackage{newtxtext,newtxmath}

\usepackage[T1]{fontenc}

\DeclareRobustCommand{\VAN}[3]{#2}
\let\VANthebibliography\thebibliography
\def\thebibliography{\DeclareRobustCommand{\VAN}[3]{##3}\VANthebibliography}

\usepackage{enumitem}

\usepackage{graphicx} 
\usepackage{amsmath} 
\usepackage{subcaption}
\usepackage{placeins}
\usepackage{orcidlink}
\usepackage[table]{xcolor}
\usepackage[nameinlink,noabbrev]{cleveref}
\creflabelformat{equation}{#2\textup{(#1)}#3}
\crefname{figure}{Fig.}{Figs.}
\crefname{table}{Table}{Tables}
\newcommand{\D}{\text{d}}

\title[Energetics of AGN Feedback]{Energetics of AGN Feedback}

\author[R. Turner et al.]{Ross J. Turner\textsuperscript{\orcidlink{0000-0002-4376-5455}},$^{1}$\thanks{Email:ross.turner@utas.edu.au}
Andrew Sullivan\textsuperscript{\orcidlink{0009-0001-0316-0304}},$^{2}$ William R. Q. Gaffney\textsuperscript{\orcidlink{0009-0002-3130-6654}}$^{1}$
\\
$^{1}$School of Natural Sciences, University of Tasmania, Private Bag 37, Hobart, 7001, Australia\\
$^{2}$International Centre for Radio Astronomy Research, The University of Western Australia, 35 Stirling Highway, Crawley, 6009, Australia}

\date{Accepted 2026 August 25. Received 2026 August 06; in original form 2026 June 29}

\pubyear{2026}

\begin{document}
\label{firstpage}
\pagerange{\pageref{firstpage}--\pageref{lastpage}}
\maketitle

\begin{abstract}
Integrating realistic active galactic nucleus (AGN) feedback into cosmological hydrodynamical simulations remains a major challenge as resolving the spatial coupling of jet energy over cosmic timescales is computationally prohibitive. We present an analytic framework that predicts the radial and polar-angle dependence of feedback energy from lobed AGNs, enabling a physically motivated, computationally efficient prescription for jet feedback. Built upon the \textit{Radio AGN in Semi-analytic Environments} (RAiSE) dynamical model, our approach tracks the post-AGN jet phase evolution of radio sources through two distinct mechanisms: buoyantly rising bubbles subject to ablation, and the gravitational collapse of swept-up gas in the shocked shell. We find that these two feedback mechanisms produce strongly contrasting spatial distributions across ten representative cluster environments. Buoyant bubbles preferentially deposit energy in steep density gradients near the core radius, while the collapsing shocked gas shell drives heating within flatter cluster cores. Weak, short-lived AGN outbursts ($Q<10^{37}$~W and $t_{\rm on}<10$~Myr) confine their energy deposition to the inner 10~kpc of the cluster; longer-lived events ($t_{\rm on} \geqslant 100$~Myr) deposit less than 1\% of their injected energy within 30~kpc. We find that time-averaged heating rates across multiple outbursts are sufficient to offset radiative cooling in all but the highest density cluster cores (within 100~kpc) for duty cycles of $0.08< \delta \leqslant 1$. This framework provides a scalable {basis for modelling anisotropic, physically motivated jet feedback that can be incorporated} into next-generation cosmological simulations.
\end{abstract}




\section{Introduction}

Hierarchical models of structure formation successfully reproduce the large-scale distribution of matter and properties of galaxies across cosmic time \citep{Press+1974,White+1978,White+1991,Lacey+1994}. Star formation shows a strong mass dependence, known as the galaxy star-formation main sequence, in which lower-mass galaxies form stars more efficiently than massive systems \citep{Bauer+2013,Chang+2015}. These models break down for galaxies in the most massive galaxy clusters: the stellar-to-halo mass ratio declines steeply \citep{Behroozi+2013} and stellar feedback becomes ineffective at regulating baryonic growth \citep{Dubois+2008,Dashyan+2018}. The intracluster medium (ICM) in cluster cores is hot, dense, and X-ray luminous \citep{Sarazin+1986,Stanek+2006}, with cooling times far below a Hubble time, implying that classical cooling flows should produce large cold gas reservoirs and sustained central star formation \citep{Fabian+1994,Hudson+2010}. This expectation is not observed: brightest cluster galaxies are instead massive, quiescent systems with low star-formation rates and long gas-depletion times \citep{Fraser+2014}. Their presence within rapidly cooling atmospheres highlights a fundamental failure of purely gravitational and stellar-feedback models, requiring an additional heating mechanism to regulate the thermal balance of cluster baryons \citep{Tamura+2001,Peterson+2003}.

To resolve this `cooling catastrophe' \citep{Cowie+1977,Fabian+1977}, {feedback from central supermassive black holes is widely invoked, with energy injected into the surrounding medium by active galactic nucleus (AGN) activity \citep{McNamara+2007,McNamara+2012,Fabian+2012}}. The efficiency and mode of this coupling depend primarily on the accretion state. At high accretion rates (i.e., $\dot{m} \geqslant 0.01$; see \citealt{Heckman+2014}), AGN operate in a radiatively efficient `quasar mode', characterised by a geometrically thin, optically thick disk \citep{Novikov+1973,Shakura+1973} that primarily drives fast, wide-angle winds capable of removing gas from the galaxy \citep{DiMatteo+2005,Fiore+2017}. By contrast, massive cluster environments are dominated by low-Eddington accretion, where hot, radiatively inefficient flows (ADAFs; \citealt{Narayan+1994}) produce collimated relativistic jets via magnetic extraction of black hole spin energy through {(e.g.)} the \citet{Blandford+1977} mechanism. These jets inflate radio lobes as they propagate through the ICM, excavating X-ray cavities observed in systems such as Perseus \citep{Boehringer+1993,Fabian+2003} and Virgo \citep{Churazov+2003,Forman+2005}. The expansion of these lobes does mechanical work on the surrounding gas, including shocks, $p\D V$ work during inflation, and work done against the cluster gravitational potential during buoyant rise \citep{Birzan+2004,McNamara+2005,Wise+2007}. In many systems, this power is sufficient to offset radiative cooling losses, maintaining approximate thermal balance in cluster cores \citep{Rafferty+2006,Hlavacek-Larrondo+2012,Russell+2013}.

Translating cluster-scale thermodynamic cycles into cosmological hydrodynamical simulations remains a significant numerical challenge. Early implementations typically adopted a single AGN feedback channel: isotropic thermal energy injection representing high-accretion quasar mode activity \citep[e.g.,][]{Schaye+2010,Hirschmann+2014,Schaye+2015,Khandai+2015,Tremmel+2017}. While this approach can regulate star formation on galaxy scales, it provides a highly averaged description of the coupling between AGN energy output and the surrounding gas. More recent simulations often use two-mode AGN feedback models in which low-accretion `kinetic mode' activity is represented by either injecting thermal or kinetic bubbles to mimic radio lobes \citep[e.g.,][]{Sijacki+2007,Vogelsberger+2014,Kaviraj+2017,Dave+2019,Dubois+2021}. 
However, the spatial scales associated with black hole accretion disks and jet launching lie many orders of magnitude below the resolution limits of cosmological simulations, necessitating the use of subgrid models to capture accretion physics and black hole spin evolution \citep[e.g.,][]{
Steinborn+2015,Fiacconi+2018,Bustamante+2019,Husko+2022}.
These subgrid approaches are not extended to their jet--environment coupling prescriptions, notably by not self-consistently capturing the evolving AGN lobe geometry in the mechanical feedback channel.

Relativistic jets and their associated lobes deposit energy in a strongly directional manner, producing elongated shock structures that uplift ambient gas {\citep{McNamara+2007,Fabian+2012,Hardcastle+2020}}, while buoyantly rising bubbles restructure the thermal profile of the gas out to large galactocentric radii {\citep[e.g.,][]{Begelman+2001,Roychowdhury+2004,Bourne+2017,English+2019}}. This leads to a highly non-spherical coupling between AGN energy and the ICM, with heating concentrated along the jet axis while orthogonal directions remain comparatively under-heated, allowing continued radiative cooling {\citep[e.g.,][]{Yang+2016,Li+2017,Martizzi+2019,Sullivan+2025}}. The efficiency of feedback therefore depends not only on the total injected energy, but {also on its spatial distribution and temporal evolution across multiple duty cycles. Capturing this behaviour has motivated numerous hydrodynamic simulations of radio jet feedback \citep[e.g., review by][]{Bourne+2023}, although their computational expense limits exploration of the broad parameter space relevant to galaxy and cluster evolution. By contrast, \citet{Raouf+2017} couple an analytic model of hemispherical radio lobes to cosmological simulations, linking lobe evolution} to the suppression of gas cooling and galaxy-scale star formation regulation. Accurately capturing this self-regulated thermodynamic cycle requires tracking the explicit radial, angular, and temporal structure of AGN energy deposition, as feedback effectiveness is inherently shaped by the dynamics of each jet outburst {\citep[e.g., self-regulated feedback; see][]{Gaspari+2012,Gaspari+2013,Gaspari+2020}}.

The limitations of current subgrid models motivate the use of the \textit{Radio AGN in Semi-analytic Environments} (RAiSE; \citealt{Turner+2015,Turner+2023a}) framework. In contrast to conventional, idealised feedback prescriptions, RAiSE explicitly tracks the expansion of radio sources in realistic cluster environments, including their evolving geometry, and the internal energy densities of the lobe and shocked gas shell. This yields a physically motivated mapping between the instantaneous jet power at the accretion disk and the spatial distribution of deposited energy on cluster scales. We introduce the RAiSE framework in \cref{sec:lobe dynamical model} and discuss modifications to consider the work done against the gravitational potential of the cluster, the internal energy of swept-up ambient gas, and bremsstrahlung cooling of dense gas in the shocked shell. In \cref{sec:feedback energetics}, we present our approach to model the radial and angular distribution of feedback energy well-after the cessation of jet activity. The feedback energetics are analysed as a function of active age and jet power for ten representative cluster environments in \cref{sec:results}. We present our conclusions in \cref{sec:conclusion} and outline methodologies for integrating our framework into cosmological simulations.

\section{Lobe Dynamical Model}
\label{sec:lobe dynamical model}

We model the radial and polar dependence of AGN feedback from radio lobes using the \textit{Radio AGN in Semi-analytic Environments} (RAiSE) model, as described by \citet{Turner+2023a} and \citet{Turner+2026}.
Their dynamical model assumes a powerful relativistic plasma jet propagating through the ambient medium, driving a bow shock that expands outwards from the jet-head. The jet plasma is deflected back towards the active nucleus by the pressure of the shocked ambient gas within the bow shock. The region filled by this shock-accelerated plasma is referred to as the lobe.

{In the following sections, we first summarise the existing RAiSE framework before extending it to account for the work done against the cluster gravitational potential, the internal energy of swept-up ambient gas, and bremsstrahlung cooling, ensuring that the injected jet energy is conserved throughout the active and remnant evolution of the radio source.}


\begin{figure*}
\includegraphics[width=0.625\textwidth,trim={0 0 0 0},clip]{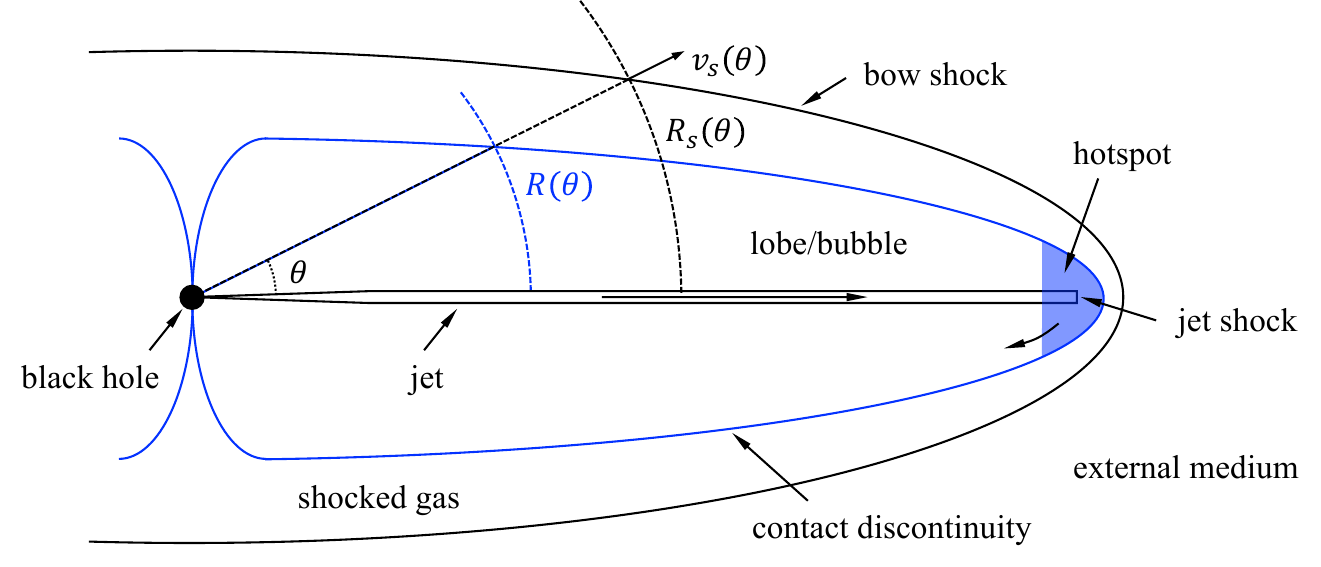}
\caption{Schematic of the \citet{Turner+2023a} dynamical model for the lobe and shocked shell (\cref{sec:pressure-driven expansion}), showing the scission of the two lobes due to fluid instability upon the cessation of jet activity \citep{Turner+2026}.}
\label{fig:schematic}
\end{figure*}

\subsection{Pressure-driven expansion}
\label{sec:pressure-driven expansion}

The expansion of the relativistic jet and shocked gas shell is modelled by \citet{Turner+2023a} considering a two-phase fluid with momentum flux contributions from both the jet and thermal shocked gas (see their Section 2). {The jet expansion phase of the RAiSE framework was developed and validated against hydrodynamic simulations spanning both light- and heavy-jet regimes. The former are modelled using a relativistic spine surrounded by a slower, denser sheath, whereas the latter are represented by a uniform non-relativistic flow \citep[][their Sections 2.2 and 4.3]{Turner+2023a}. These two jet prescriptions lead to different large-scale evolutionary histories, captured by the analytic model. The present work adopts their light-jet model, representative of the powerful relativistic jets responsible for extended \citet{Fanaroff+1974} Type-II radio sources \citep[e.g.,][]{Hardcastle+2020}.}

We focus on their equations describing the later-time evolution of the shocked shell in this summary (closely following \citealt{Turner+2026}){; the jet expansion phase (above) provides the initial conditions for this subsequent evolution}. {The governing equations are derived from conservation of momentum for the expanding shocked shell \citep[][their Equation 17]{Turner+2023a} and the first law of thermodynamics (their Equation 15), assuming pressure equilibrium is maintained between the lobe and shocked shell. The injected jet power is partitioned between increasing the internal energy of this lobe--shocked shell system and performing work during adiabatic expansion.}

The RAiSE framework solves a system of ordinary differential equations (ODEs; derivatives with respect to the source age, $t$) for the radius, velocity and Lorentz factor at the surface of the shocked shell as a function of the polar angle $\theta$ from the jet axis, i.e., $R_s(t, \theta)$, $v_s(t, \theta)$ and $\gamma_s(t, \theta)$, respectively. \citet{Turner+2023a} adopt a numerical scheme using a fourth-order Runge-Kutta method in terms of a system of three non-linear first order ODEs for each polar angle.
The following system of equations must be solved for each small angular element $[\theta - \delta\theta/2, \theta + \delta\theta/2)$ of the shocked shell \citep[see \cref{fig:schematic} and][their Equation 19]{Turner+2023a}:
\begin{equation}
\begin{split}
\dot{R}_\text{s}(t, \theta) &= v_\text{s} \\
\dot{v}_\text{s}(t, \theta) &= \frac{3\Gamma_{\rm c} - \beta}{2 R_\text{s} [1 + (\gamma_\text{s} v_\text{s}/c)^2] [\zeta_{\rm s} \gamma_\text{s}/\eta_{\rm s}]^2} \\
&\quad\quad \times 
\bigg[\frac{3 (\Gamma_{\rm c} - 1) Q R_s^{\!\;\beta - 2} \eta_{\rm s}^{3 - \beta}\zeta_{\rm s}^{2}}{2 \pi (3\Gamma_{\rm c} - \beta)I\!\!\!\;\; v_\text{s} (\rho a^\beta)} - \Big(\frac{\zeta_{\rm s} \gamma_\text{s} v_\text{s}}{\eta_{\rm s}}\Big)^2 - \frac{c_{\rm x}^2}{\Gamma_{\rm c}} \bigg]
\\
\dot{\gamma}_\text{s}(t, \theta) &= \frac{\gamma_\text{s}^3 v_\text{s} \dot{v}_\text{s}}{c^2} ,
\end{split}
\label{supersonic system}
\end{equation}
where $Q$ is the instantaneous jet power (each jet; i.e., $Q_\text{tot} = 2Q$), $c$ is the speed of light, $c_{\rm x}$ is the sound speed of the ambient medium, and $\Gamma_c = \tfrac{5}{3}$ is the adiabatic index of the lobe plasma and shocked gas. The ambient gas density is locally described by a power law of the form:
\begin{equation}
\rho_\text{x}(r) = \rho [r/a]^{-\beta} ,
\end{equation}
{where $\rho$ is the density at some arbitrary scale radius $a$ (e.g., the core density and core radius).}
The RAiSE framework models complex gas density profiles as a continuous function that combines numerous such power laws each defined over small radial intervals. The variable $I$ is a source-specific constant that represents the steady state distribution of energy throughout the lobe and shocked shell by the jet. That is,
\begin{equation}
I = \int_0^{\tfrac{\pi}{2}} {\eta_{\rm s}}^{3 - \beta}(\theta)\!\; {\zeta_{\rm s}}^2(\theta) \sin\theta \D\theta ,
\label{I integral}
\end{equation}
where $\eta_s$ and $\zeta_s$ are geometric factors describing the initially ellipsoidal shape of the shocked shell defined in Equations 12 and 13 of \citet{Turner+2023a}, respectively. 

\subsection{Gravitational-driven suppression}
\label{sec:gravity}

{We first extend the RAiSE framework to account for the work done against the gravitational potential of the cluster.} The displacement of dense ambient gas by the expanding lobe out to the higher radii of the shocked gas shell does a not insignificant amount of work against {this potential}.

The local gravitational acceleration, $g(r)$, acting on the ambient gas
is related to the equilibrium gas density profile, $\rho_\text{x}(r)$, as follows \citep[e.g.,][their equation 14]{Turner+2026}:
\begin{equation}
g(r) = -\frac{1}{\rho_\text{x}(r)} \frac{\D p_\text{x}(r)}{\D r} \quad = \frac{\beta(r) c_\text{x}^2}{r \Gamma_\text{c}},
\label{gravity}
\end{equation}
where $p_\text{x}(r)$ is the ambient pressure profile, related to the density under the assumption of an approximately isothermal gas. The gravitational potential is locally approximated as the integral of \cref{gravity}, giving,
\begin{equation}
\Phi(r) = \frac{\beta(r) c_\text{x}^2}{\Gamma_\text{c}} \log r + \Phi_0,
\label{potential}
\end{equation}
where $\Phi_0$ is a local reference value of the potential.

The energy lost by each angular element of the lobe--shocked shell system (i.e., $\delta U_\text{grav}(\theta) < 0$) due to work done against the gravitational potential of the cluster is given by \citep[cf.][their Section 4.1]{Turner+2026},
\begin{equation}
\begin{split}
&\delta U_\text{grav}(t, \theta) = 2\pi\, \bigg[ \int_0^{R(\theta)} \!\!\rho_\text{x}(r) \Phi(r) r^2 \D r \\
&\qquad + \int_{R(\theta)}^{{R_\text{s}(\theta)}} \!\!\rho_\text{x}(r) [1 - \varrho_\text{s}(\theta)] \Phi(r) r^2 \D r \bigg] \sin\theta \delta \theta .
\end{split}
\label{work}
\end{equation}
We compare the mass of ambient plasma swept up into each angular element of the dense shocked gas shell to the lower mass in that region for gas in the hydrostatic equilibrium state; i.e., the overdensity of the shocked gas shell, $\varrho_\text{s}(\theta)$. This overdensity is thus given by the ratio \citep[cf.][their equations 16 and 17]{Turner+2026},
\begin{equation}
\begin{split}
\varrho_\text{s}(t, \theta) &= \int_0^{R_\text{s}(\theta)} \!\!\rho_\text{x}(r) r^2 \D r \,\Big/ \int_{R(\theta)}^{R_\text{s}(\theta)} \!\!\rho_\text{x}(r) r^2 \D r \\
&= \frac{(3 - \beta) R_\text{s}^{\beta - 3} m_\text{s}}{\rho a^\beta [1 - (R_\text{s}/R)^{\beta - 3}]} \quad \approx \frac{1}{1 - (R_\text{s}/R)^{\beta - 3}},
\end{split}
\label{overdensity}
\end{equation}
where the second equality is obtained by locally approximating $\rho_\text{x}(r)$ as a power law between $R(\theta)$ and $R_\text{s}(\theta)$; the final approximate form applies for a single power law profile over the entire ambient medium, and is provided solely to aid in interpretation of the overdensity.
The mass of the shocked shell in the former general expression is calculated numerically by solving an additional equation in the system of ODEs. That is, 
\begin{equation}
\begin{split}
\dot{m}_\text{s}(t, \theta) = \rho a^\beta R_\text{s}^{2-\beta} v_\text{s} ,
\end{split}
\label{fourth equation}
\end{equation}
where the $\sin\theta \delta \theta$ term cancels in subsequent equations so is neglected in this definition for clarity. 

The lobe and shocked gas shell are known to expand self-similarly upon the formation of the lobe \citep[e.g.,][]{Falle+1991,KA+1997}, at least for mildly supersonic sources \citep{Turner+2015}. The ratio of the shocked shell to lobe radii along the jet axis is taken as $b \approx 1.07${, as determined by \citet{Turner+2023a} from hydrodynamic simulations.} This ratio varies with polar angle as $R_\text{s}(\theta)/R(\theta) = b\eta_\text{s}(\theta)/\eta(\theta)$ for the geometric factors $\eta$ and $\eta_\text{s}$ defined by \citet[][their equations 12 and 14]{Turner+2023a}. The overdensity of the shocked gas shell is therefore approximately time independent.

The differential equations in the RAiSE framework are expressed as derivatives of key parameters of the system. We therefore derive an energy loss rate as follows:
\begin{equation}
\begin{split}
&\frac{\D[\delta U_\text{grav}(t,\theta)]}{\D t} = 2\pi\, \Big[ \rho_\text{x}(R) \varrho_\text{s} \Phi(R) R^2 \frac{\D R}{\D t} \\
&\qquad + \rho_\text{x}(R_\text{s}) [1 - \varrho_\text{s}] \Phi(R_\text{s}) R_\text{s}^2 \frac{\D R_\text{s}}{\D t} \Big] \sin\theta \delta \theta ,
\end{split}
\end{equation}
where we assume the overdensity, $\varrho_\text{s}$, is a slowly varying function of time.
We can greatly simplify this expression by choosing $\Phi_0$ such that $\Phi(R) = 0$, and applying \cref{overdensity} for the overdensity of the shocked gas shell. That is,
\begin{equation}
\begin{split}
\frac{\D[\delta U_\text{grav} (t,\theta)]}{\D t} &= \frac{2\pi (\rho a^\beta) [1 - \varrho_\text{s} ]\beta c_\text{x}^2 R_\text{s}^{2 - \beta} v_\text{s}}{\Gamma_\text{c}} \log(b \eta_\text{s}/\eta ) \sin\theta \delta \theta \\
&\approx \frac{2\pi (\rho a^\beta) \beta c_\text{x}^2 R_\text{s}^{2 - \beta} v_\text{s}}{\Gamma_\text{c}[1 - (b\eta_\text{s}/\eta)^{3 - \beta}]} \log(b \eta_\text{s}/\eta ) \sin\theta \delta \theta ,
\end{split}
\label{gravity de}
\end{equation}
where energy is lost from the system (i.e., $\delta U_\text{grav}(\theta) < 0$) if $v_\text{s} > 0$ (i.e., shell expanding outwards) and gains energy if the shell is collapsing inwards \citep[e.g.,][]{Turner+2026}. Importantly, deviations to the ratio $b \eta_\text{s}/\eta \equiv 1 + \epsilon$ from the chosen self-similar value are unimportant as $\log (1 + \epsilon)/[1 - (1 + \epsilon)^{3 - \beta}] = -1/(3 - \beta) + \tfrac{1}{2}\epsilon + \mathcal{O}(\epsilon^2)$.

The expansion of the shocked shell is found using the following relationship for adiabatic expansion of a small angular volume element with momentum flux $p_\text{s}(\theta)$ and volume $\delta V_\text{s}(\theta)$ \citep[cf.][their equation 15]{Turner+2023a}:
\begin{equation}
\begin{split}
&\frac{\D p_\text{s}(t,\theta)}{\D t} \delta V_\text{s}(t,\theta) + \Gamma_\text{c} p_\text{s}(t,\theta) \frac{\D [\delta V_\text{s}(t,\theta)]}{\D t} \\
&\qquad = (\Gamma_\text{c} - 1) \Big[Q \delta \lambda(\theta) + \frac{\D[\delta U(t,\theta)]}{\D t} \Big],
\label{first law grav}
\end{split}
\end{equation}
where $\delta \lambda(\theta) = \eta_\text{s}^{3-\beta} \zeta_\text{s}^2 \sin\theta \delta \theta /I$ (see \cref{I integral}) is the fraction of that power associated with the expansion of the volume $\delta V_\text{s}(\theta)$.

The work done by the lobe--shocked shell system against the gravitational potential of the cluster can be added into the system of ODEs derived by \citet{Turner+2023a} as follows:
\begin{equation}
\begin{split}
\dot{v}_\text{s}(t, \theta) &= \frac{3\Gamma_{\rm c} - \beta}{2 R_\text{s} [1 + (\gamma_\text{s} v_\text{s}/c)^2] [\zeta_{\rm s} \gamma_\text{s}/\eta_{\rm s}]^2} \\
&\quad\quad \times \bigg[\frac{3 (\Gamma_{\rm c} - 1) Q R_s^{\!\;\beta - 2} \eta_{\rm s}^{3 - \beta}\zeta_{\rm s}^{2}}{2 \pi (3\Gamma_{\rm c} - \beta)I\!\!\!\;\; v_\text{s} (\rho a^\beta)} - \Big(\frac{\zeta_{\rm s} \gamma_\text{s} v_\text{s}}{\eta_{\rm s}}\Big)^2 - \frac{(1 + \Delta)c_{\rm x}^2}{\Gamma_{\rm c}} \bigg],
\end{split}
\label{supersonic system grav}
\end{equation}
where we define a gravitational correction to the sound speed term of the form:
\begin{equation}
\begin{split}
\Delta_\text{grav}(t, \theta) &= \frac{3 \beta (\Gamma_{\rm c} - 1) [\varrho_\text{s} - 1]\log(b \eta_\text{s}/\eta )}{(3\Gamma_{\rm c} - \beta)} \\
&\approx \frac{3\beta(\Gamma_\text{c} - 1)}{(3\Gamma_\text{c}  - \beta)(3 - \beta)} ,
\end{split}
\label{Delta 1}
\end{equation}
where the second equality assumes the previous Taylor series approximation for the ratio $b \eta_\text{s}/\eta \equiv 1 + \epsilon$. This correction is bounded such that $\Delta_\text{grav} \in [0, \tfrac{4}{3}]$ for $\Gamma_\text{c} = \tfrac{5}{3}$ and $0 \leqslant \beta \leqslant 2$. Consequently, the gravitational potential will have negligible effect on the dynamics of the lobe in flatter sections of the ambient medium (i.e., $\Delta_\text{grav} \approx 0$ {when $\beta \approx 0$}), but will slow their expansion in steeper regions {($\beta \gg 0$)} towards the outskirts of the cluster with $\Delta_\text{grav} \gtrsim 1$.

%
%

\subsection{Swept-up gas internal energy}
\label{sec:swept-up gas internal energy}

The ambient medium swept-up by the expanding lobe and compressed in the shocked gas shell comprises a non-negligible internal energy that remains within the lobe--shocked shell system{; we therefore extend the RAiSE framework to include this contribution.}

The mass of ambient gas in the shocked shell is derived numerically by solving our added fourth differential equation in the RAiSE {model} (i.e., \cref{fourth equation}). The rate of change in the internal energy of swept-up ambient gas is related to the derivative of this mass as follows:
\begin{equation}
\begin{split}
&\frac{\D[\delta U_\text{int}(t,\theta)]}{\D t} = \frac{2\pi \dot{m}_\text{s} c^2_\text{x}}{\Gamma_\text{c}(\Gamma_\text{c} - 1)} \sin\theta \delta \theta.
\end{split}
\label{internal swept}
\end{equation}
The internal energy of swept-up ambient gas consumed by the lobe--shocked shell system can be included in the system of ODEs as for the work against the gravitational potential. We define an ambient internal energy correction to the sound speed term in \cref{supersonic system grav} of the form:
\begin{equation}
\begin{split}
\Delta_\text{int}(t, \theta) &= -\frac{3 R_\text{s}^{\beta - 2} \dot{m}_\text{s}}{(3\Gamma_\text{c} - \beta) v_\text{s}(\rho a^\beta)} \quad = -\frac{3}{3\Gamma_\text{c}  - \beta} ,
\end{split}
\label{Delta 2}
\end{equation}
where the second equality is obtained using \cref{fourth equation}.

The sum of the gravitational and ambient internal energy corrections (i.e., $\Delta_\text{grav} + \Delta_\text{int}$) yields:
\begin{equation}
\begin{split}
\Delta(t, \theta) 
&= \frac{3 \beta (\Gamma_{\rm c} - 1) [\varrho_\text{s} - 1]\log(b \eta_\text{s}/\eta ) - 3}{(3\Gamma_{\rm c} - \beta)} \\
&\approx \frac{3(\beta\Gamma_\text{c} - 3)}{(3\Gamma_\text{c}  - \beta)(3 - \beta)} ,
\end{split}
\label{Delta 3}
\end{equation}
where the correction is bounded such that $\Delta \in [-\tfrac{3}{5}, \tfrac{1}{3}]$ for $\Gamma_\text{c} = \tfrac{5}{3}$ and $0 \leqslant \beta \leqslant 2$. The internal energy of the swept-up ambient gas provides a boost to the expansion rate in flatter sections of the ambient medium (i.e., $\Delta < 0$) but is overcome by the relatively large energy drain of the gravitational potential in steeper regions towards the outskirts of the cluster. This correction only applies for outward expansion as, in particular, the consumption of internal energy from the ambient gas is not a reversible process (e.g., $\Delta \approx \Delta_\text{grav}$ for $v_\text{s} < 0$). 

The significance of the gravitational potential and the internal energy of the swept-up ambient gas on the expansion and evolution of the lobe--shocked shell system is investigated in \cref{sec:results}.

\subsection{Bremsstrahlung X-ray cooling}
\label{sec:Bremsstrahlung cooling}

We consider the energy loss rate due to bremsstrahlung cooling of the dense gas in the shocked shell. The volumetric cooling rate in the shocked gas shell integrated across X-ray wavelengths is approximated as follows \citep[e.g.,][]{1979rpa..book.....R}:
\begin{equation}
    \mathcal{C}(t, r, \theta) = \Big(\frac{3\rho_\text{s}(r)}{4 m_\text{p}}\Big)^2 \Lambda_0 \tau_\text{s}^{1/2}(r) \quad = \frac{9\Lambda_0}{16} \Big(\frac{p_\text{s}(R_\text{s}) \rho^3_\text{s}(r) \bar{\mu}}{k_\text{b} m^3_\text{p}} \Big)^{1/2},
    \label{cooling rate}
\end{equation}
where $\Lambda_0 = 1.43 \times 10^{-40}$~W\,m$^{3}$\,K$^{-1/2}$ and $m_\text{p}$ is the proton mass. The constants in the second equality are the dimensionless mean molecular weight, $\bar{\mu} = 0.6$, and the Boltzmann constant, $k_\text{b}$. 

The energy loss rate due to cooling (i.e., $\delta U_\text{cool} < 0$) of a given angular element of the shocked gas shell is given by,
\begin{equation}
\begin{split}
\frac{\D[\delta U_\text{cool}(t,\theta)]}{\D t} &= -2\pi \int_{R(\theta)}^{R_\text{s}(\theta)} \mathcal{C}(t, r, \theta) r^2 \D r \sin\theta \delta \theta \\
&= -\frac{3\pi \Lambda_0 R_\text{s}^{3(1 - \beta/2)} [1 - (R_\text{s}/R)^{3(\beta/2 -1)}]}{4(2 - \beta)} \\
&\qquad\quad \times \Big(\frac{p_\text{s}(R_\text{s}) [\varrho_\text{s} \rho a^\beta]^3 \bar{\mu}}{k_\text{b} m^3_\text{p}} \Big)^{1/2} \sin\theta \delta \theta ,
\end{split}
\label{cooling energy}
\end{equation}
where the pressure of the shocked shell is defined by \citet[][their equation 17]{Turner+2023a}. We assume the density of the shocked gas shell is uniformly increased by a factor of the overdensity, $\varrho_\text{s}$ (\cref{overdensity}), relative to the equilibrium density at a given location. The cooling rate shares minimal resemblance to any of the terms in the system of ODEs derived by \citet{Turner+2023a}, however, following the approach of the previous two sections, we define a cooling correction, $\Delta_\text{cool}$, to the sound speed term. That is,
\begin{equation}
\begin{split}
\Delta_\text{cool}(t, \theta) &= \frac{9\Gamma_\text{c}(\Gamma_\text{c} - 1)\Lambda_0 (\rho a^\beta) R_\text{s}^{1 - \beta} [1 - (R_\text{s}/R)^{3(\beta/2 -1)}]}{8(2 - \beta)(3\Gamma_\text{c} - \beta) v_\text{s} c^2_\text{x}} \\
&\qquad\quad \times \Big(\frac{\varrho^3_\text{s} \bar{\mu}}{k_\text{b} m^3_\text{p}} \Big)^{1/2\,} \bigg[\Big(\frac{\zeta_{\rm s} \gamma_\text{s} v_\text{s}}{\eta_{\rm s}}\Big)^2 + \frac{c_{\rm x}^2}{\Gamma_{\rm c}} \bigg]^{1/2} ,
\end{split}
\label{Delta 4}
\end{equation}
where this cooling correction adds as $\Delta = \Delta_\text{grav} + \Delta_\text{int} + \Delta_\text{cool}$.
%
%
%
%

We briefly discuss the relative importance of bremsstrahlung cooling on the dynamics of the lobe--shocked shell system, referring to the evolutionary tracks considered in \cref{sec:results}. The cooling correction is initially zero, reaching $\Delta_\text{cool} = 0.003$-$0.005$ at 300~Myr for our clusters with lower core densities (e.g., left panel of \cref{fig:dynamics}) and $\Delta_\text{cool} = 0.007$-$0.043$ in clusters with higher densities (e.g., centre and right panels). We require significantly denser environments or older sources for bremsstrahlung cooling to have a non-negligible effect on the dynamics.

\section{Feedback Energetics}
\label{sec:feedback energetics}

We analyse the state of the cluster ambient medium during and upon the cessation of jet activity to determine the coupling efficiency of AGN feedback. The heating rate of the lobe and shocked gas shell during the active and remnant phase are considered in \cref{sec:shocked shell and lobe}. We model the heating at large galactocentric radii due to buoyantly rising remnant bubbles for both adiabatic expansion and ablation of lobe plasma through fluid instability or turbulent motion (\cref{sec:bubbles}).

\subsection{Shocked shell and lobe}
\label{sec:shocked shell and lobe}

The dynamics of the lobe--shocked shell system, described by the system of ODEs in \cref{sec:lobe dynamical model}, self-consistently converts between different forms of energy during the active phase; i.e., the heating rate of the AGN  at a given location can be directly calculated from the change in internal energy of the plasma relative to its equilibrium state. This change in internal energy implicitly includes the contributions from (e.g.) shock heating and turbulent motion; the former is directly captured in the conservation equations underpinning the RAiSE framework, the latter is included in the internal energy as small-scale fluid flows are not resolved in a kinetic energy term.


The change in internal energy density of the lobe--shocked shell system during the active phase at some radius and polar angle is given by \citep[][their equation 17]{Turner+2023a},
\begin{equation}
\begin{split}
&u_\text{int}(t, r, \theta) = \bigg[\frac{\rho_\text{x}(R_\text{s})}{\Gamma_\text{c} - 1}\Big(\frac{\zeta_{\rm s} \gamma_\text{s} v_\text{s}}{\eta_{\rm s}}\Big)^2 + \frac{(\rho_\text{x}(R_\text{s}) - \rho_\text{x}(r))c_{\rm x}^2}{\Gamma_{\rm c}(\Gamma_\text{c} - 1)} \bigg] \\
&\qquad \times [1 - H(r - R_\text{s}(\theta))] ,
\end{split}
\label{shell energy}
\end{equation}
where $H(x)$ is the Heaviside unit step function. This calculation is performed during the active phase of the jet directly using the RAiSE model outputs (see \cref{sec:lobe dynamical model}).


Upon the cessation of jet activity, however, the underdense lobe may either push through the shocked gas shell, rising buoyantly to large radii (see \cref{sec:bubbles}), {or, for sufficiently weak and old jets in dense environments, collapse inwards due to an impulsive pressure imbalance or fluid instabilities, depending on the magnetic field strength \citep[for a complete description, see][]{Turner+2026}}. The dense shocked gas shell (in either case) will collapse towards the centre of the gravitational potential of the cluster.

We consider the state of the shocked gas upon returning to an approximate hydrostatic equilibrium (primarily through mixing, i.e., no work done), albeit with some small increase in internal energy. The increase is assumed be be distributed in proportion to the equilibrium internal energy density of the ambient gas within the region occupied by the lobe--shocked shell system at the cessation of jet activity. The change in internal energy density of the ambient gas at some radius and polar angle is therefore related to the energy density at the end of the active phase, $t_\text{on}$ (cf. \cref{shell energy}), as follows:
\begin{equation}
\begin{split}
&u_\text{int}(t \gg t_\text{on}, r, \theta) = \\
&\qquad \rho_\text{x}(r) [1 - H(r - R_\text{s}(t_\text{on},\theta))] \, \Big / \int_0^{R_\text{s}(t_\text{on},\theta)} \! \rho_\text{x}(r) \, r^2\D r \\
&\qquad\qquad \times \Big( \int_{r'}^{R_\text{s}(t_\text{on},\theta)} u_\text{int}(t_\text{on}, r, \theta)  \, r^2 \D r + \frac{\delta U_\text{grav}(t_\text{on}, \theta)}{\sin\theta \delta \theta} \Big),
\end{split}
\label{ambient energy}
\end{equation}
where the lower bound of the second integral is $r' = R(t_\text{on},\theta)$ if the lobe rises buoyantly to large radii, and $r' = 0$ if the lobe collapses inwards along with the shocked gas.
The change in gravitational potential energy of the lobe--shocked shell system along some polar angle, $\delta U_\text{grav}(t_\text{on}, \theta)$, is found by integrating \cref{gravity de} in the RAiSE framework as for the ODEs describing the dynamics of the system.

This approach assumes the shocked shell does not expand upon entering the remnant phase, which is unlikely for high-Mach number shocks \citep[cf.][]{Kaiser+2002}. We assess the sensitivity of our results to continued growth (or `coasting') at early-times in the remnant phase using the \citet{Turner+2018} dynamical model (see discussion in \cref{sec:results}). However, their model remains valid only until the buoyant bubble punctures the shocked shell. We do not explicitly include this process in our work due to significant uncertainties in the relevant fluid instability timescales. {This coasting model nevertheless captures the early supersonic stage of the subsequent propagation before the expansion velocity falls below the sound speed and the shocks broaden into sound waves \citep[e.g.,][]{Fabian+2003}. The subsequent propagation and dissipation of these waves are not explicitly modelled and may redistribute the deposited energy to larger radii and over a broader range of polar angles than predicted by the present model. However, because this redistribution occurs over a much larger solid angle than the buoyantly rising radio bubbles modelled in this work (see \cref{sec:bubbles}), the resulting volumetric heating rates are expected to be substantially lower.}

\subsection{Buoyant bubbles}
\label{sec:bubbles}

The heating rate of the buoyantly rising bubble is modelled following \citet{Begelman+2001}, \citet{Roychowdhury+2004} and \citet{Sullivan+2026}. These authors assume the bubble releases gravitational potential energy into the ambient medium as gas displaced by the bubble falls back towards equilibrium; the ambient medium gains energy in the form of bulk or turbulent motion, however, changes to the pressure are assumed to be small relative to those of the hydrostatic equilibrium profile \citep[cf.][]{Sullivan+2025}. 


The general expression for the adiabatic expansion of a remnant bubble is given by \citep[cf.][their equation 15]{Turner+2023a}:
\begin{equation}
\frac{\D p(r)}{\D r} V(r) + \Gamma_\text{c} p(r) \frac{\D V(r)}{\D r} + \Gamma_\text{c} \alpha p(r) A(r) = 0,
\label{first law}
\end{equation}
where $\Gamma_\text{c}$ is the adiabatic index of non-relativistic plasma, and $\alpha > 0$ is the spatial ablation rate at the rear of the lobe with cross-sectional area $A(r)$. For a spherical lobe, the cross-sectional area is related to the volume as $A(r) = \pi[3V(r)/4\pi]^{2/3}$; we will absorb the constants of order unity into the ablation rate for clarity.

The volume of an adiabatically expanding bubble can be related to its pressure as $V(r) = C \{p(r)\}^{-1/\Gamma_\text{c}}$, for some constant $C$ encoding to the conditions of the lobe at the cessation of jet activity. We modify this relationship to consider the ablation of the bubble by assuming the ambient pressure profile is locally approximated by a power law of the form $p_\text{x}(r) \propto r^{-\beta}$ (i.e., isothermal profile). We can therefore write \cref{first law} as a first-order differential equation in the volume as follows:
\begin{equation}
\frac{\D V(r)}{\D r} - \frac{\beta}{r \Gamma_\text{c}} V(r) + \alpha V^{2/3}(r) = 0 .
\label{first law}
\end{equation}
The solution to this equation gives an approximate form for the evolution of the bubble volume,
\begin{equation}
\begin{split}
V(r) = \Big( C^{1/3} \{p(r)\}^{-1/(3\Gamma_\text{c})} - \frac{r \Gamma_\text{c} \alpha}{3\Gamma_\text{c} - \beta} \Big)^3 ,
\end{split}
\label{bubble adiab}
\end{equation}
which converges to the standard adiabatic expression when the spatial ablation rate is $\alpha = 0$.
This expression equals zero at the following radius (for the local power law approximation):
\begin{equation}
R_\text{max} = R_\text{s} \Big( \frac{C^{1/3} (3\Gamma_\text{c} - \beta)}{R_\text{s} \Gamma_\text{c} \alpha } \{p(R_\text{s})\}^{-1/(3\Gamma_\text{c})} \Big)^{\frac{3\Gamma_\text{c}}{3\Gamma_\text{c} - \beta}} .
\label{Rmax}
\end{equation}
where this radius approaches $R_\text{max} \rightarrow \infty$ in the limit $\alpha \rightarrow 0$; i.e., the volume of the bubble remains finite if there is no ablation.

We consider the work done by the buoyant bubble due to this `adiabatic' expansion and against the gravitational potential of the cluster. The work done per unit length, $\D r$, of ambient medium at galactocentric radius, $r$, is given by,
\begin{equation}
\begin{split}
\tilde{u}(r) &= p(r) \frac{\D V(r)}{\D r} + \alpha p(r) A(r) + \rho_\text{x}(r) V(r) g(r) \\
&= p(r) \frac{\D V(r)}{\D r} + \alpha p(r) V^{2/3}(r) - V(r) \frac{\D p(r)}{\D r}, 
\end{split}
\label{u tilde 1}
\end{equation}
where we express the local gravitational potential in terms of the hydrostatic equilibrium ambient pressure profile (\cref{gravity}). This expression is further simplified by removing the explicit volume dependence using \cref{bubble adiab}. The work done per unit length is then given by,
\begin{equation}
\begin{split}
&\tilde{u}(r) = p(r) \Big( C^{1/3} \{p(r)\}^{-1/(3\Gamma_\text{c})} - \frac{r \Gamma_\text{c} \alpha}{3\Gamma_\text{c} - \beta} \Big)^2 \\
&\qquad \times \Big( \frac{C^{1/3}(\Gamma_\text{c} + 1)\beta}{r \Gamma_\text{c}} \{p(r)\}^{-1/(3\Gamma_\text{c})} - \frac{(\Gamma_\text{c} + 1)\alpha \beta}{3\Gamma_\text{c} - \beta} \Big) ,
\end{split}
\label{u tilde 2}
\end{equation}
where we assume the pressure of the bubble approximately follows the hydrostatic equilibrium profile. The constant $C$ is related to the conditions of the lobe at the cessation of jet activity such that $C = p^{1/\Gamma_\text{c}}(R_\text{s})V(R_\text{s})$.

%
%
%
%
The total work done by the buoyant bubble can be solved analytically for $\alpha =0$ by integrating radially outwards from the initial radius (e.g., minor axis of the shocked gas shell; cf. \citealt{Sullivan+2026}). That is,
\begin{equation}
\begin{split}
U_\text{lobe} &= \int_{R_\text{s}}^\infty -\frac{C(\Gamma_\text{c} + 1)}{\Gamma_\text{c} - 1} \frac{\D}{\D r} \Big(\{p(r)\}^{(\Gamma_\text{c} - 1)/\Gamma_\text{c}} \Big) \,\D r \\
&= \frac{\Gamma_\text{c} + 1}{\Gamma_\text{c} - 1} p(R_\text{s})V(R_\text{s}),
\end{split}
\label{u lobe}
\end{equation}
where we take $p(r) \rightarrow 0$ as $r \rightarrow \infty$. The buoyant bubble is therefore expected to deposit the $4p\D V$ work involved in creating the lobe as it rises out of the cluster potential. We assess the importance of ablation on the radial distribution of AGN feedback energetics in \cref{sec:results}.

The polar-angle dependence of the work done by the buoyant bubble is modelled based on the solid angle occupied by the bubble as a function of radius. That is,
\begin{equation}
\begin{split}
\Omega_\text{lobe}(r) = 2\pi[1 - \cos\theta_\text{lobe}(r)],
\end{split}
\end{equation}
where the maximum polar angle of the bubble is approximated by the cross-sectional area $A(r)$ as follows:
\begin{equation}
\begin{split}
\theta_\text{lobe}(r) = \arccos\Big(1 - \frac{A(r)}{2\pi^2 r^2}\Big).
\end{split}
\label{crit angle}
\end{equation}
The cross-sectional area is matched to the RAiSE prediction at the cessation of jet activity and evolved as $A(r) \propto V^{2/3}(r)$ following \cref{bubble adiab}. The change in energy density of the ambient medium at some radius and polar angle due to the work done by the buoyant bubble is therefore given by, 
\begin{equation}
\begin{split}
u(t \gg t_\text{on}, r, \theta) = \frac{\tilde{u}(r)}{r^2 \Omega_\text{lobe}(r)} \begin{cases} 1, &0 \leqslant \theta \leqslant \theta_\text{lobe}(r) \\
0, &\theta > \theta_\text{lobe}(r)
\end{cases},
\end{split}
\end{equation}
where we assume the buoyant bubble provides no additional heating within the confines of the shocked shell. The calculations in this work make use of this complete expression, however, we will frequently refer to the simpler \cref{u tilde 2} in the interpretation of results.

\section{Results}
\label{sec:results}

\subsection{Lobe--shocked shell dynamics}
\label{sec:lobe--shocked shell dynamics}

{We compare the dynamics of our improved RAiSE model (see \cref{sec:lobe dynamical model}) with both our previous version of the model and the PLUTO hydrodynamic simulation \citep{Mignone+2007} originally used to calibrate the RAiSE framework \citep{Turner+2023a}. This comparison tests that the updated formulation preserves the existing calibration against hydrodynamic simulations. We further} discuss the properties of a set of mock cluster environments in \cref{sec:Cluster environments} and investigate the analytic model predictions as a function of source age, jet power and ambient medium in \cref{sec:Gravity and internal energy}.

\subsubsection{Cluster environments}
\label{sec:Cluster environments}

We consider a representative sample of mock cluster environments in hydrostatic equilibrium. The RAiSE framework can consider general, but spherically-symmetric, ambient gas density profiles with an isothermal temperature profile (cf. $\tau_\text{x} \propto lr^{-\xi}$ in \citealt{Turner+2015}); we present only `simple' environments in this work to aid interpretation of results. 

\begin{table*}
\begin{center}
\newcolumntype{L}{>{\raggedright\arraybackslash}m{100pt}}
\caption[]{Properties of mock clusters considered in this work: cluster name (first), core density, core radius, and slope at large galactocentric radii (see \cref{beta profile}; second, third and fourth), gas temperature (fifth), virial mass and gas fraction (sixth and seventh).}
\label{tab:clusters}
\renewcommand{\arraystretch}{1.1}
\setlength{\tabcolsep}{6pt}
\begin{tabular}{cccccccL}
\hline\hline
Cluster&Core Density, $\rho_\text{c}$&Core Radius, $r_\text{c}$&$\beta'$&Temperature, $\tau_\text{x}$&Virial Mass, $M_{200}$&$f_\text{gas,200}$&Comments \\
&($\times 10^{-24}$~kg\,m$^{-3}$)&(kpc)&&($\times 10^7$~K)&($\times 10^{14}$~M$_\odot$)\\
\hline
A&2.41&144&0.38&3.46&1.34&0.216&Ambient medium of reference hydrodynamic simulation.\\
B & 2.41  & 72  & 0.38 & 1.73 & 0.479 & 0.148 & \\
C & 4.82  & 72  & 0.38 & 6.92 & 3.87 & 0.137 & \\
\hline
D & 14.5  & 28.8  & 0.57 & 0.692 & 0.225 & 0.106 & \\
E & 7.23  & 72  & 0.57 & 1.73 & 0.884 & 0.113 & \\
F & 14.5  & 72  & 0.57 & 3.46 & 2.51& 0.128 & \\
G & 4.82  & 144 & 0.57 & 3.46 & 2.48& 0.132 & \\
\hline
H & 72.3  & 28.8  & 0.76 & 0.692 & 0.346 & 0.109 & \\
J & 14.5  & 72  & 0.76 & 0.692 & 0.342 & 0.153 & \\
K & 24.1  & 72  & 0.76 & 1.73 & 1.36 & 0.098 & \\
\hline
\end{tabular}
\end{center}
\end{table*}

The gas density profile of the ambient medium (in this work) is therefore modelled using a modified $\beta$-model \citep[e.g.,][]{Sarazin+1986} of the form:
\begin{equation}
\rho_\text{x}(r) = \rho_\text{c} \big( 1 + [r/r_\text{c}]^2 \big)^{-3\beta'/2} ,
\label{beta profile}
\end{equation}
where $\rho_\text{c} = 2.41 \times 10^{-24}$~kg\,m$^{-3}$, $r_\text{c} = 144$~kpc and $\beta' = 0.38$ in our reference hydrodynamic simulation from \citet{Yates+2022}; see \citet[][their section 4.1]{Turner+2023a} for a description of relevant parameters. The reference hydrodynamic simulation has an isothermal environment with a temperature of $\tau_\text{x} = 3.46\times 10^7$~K.

The virial mass of the isothermal cluster environment is found by equating the standard spherical hydrostatic equation \citep[e.g.,][]{Sarazin+1986} for the modified $\beta$-model to the mass of a sphere with an average density a factor of 200 greater than the critical density of the universe, $\rho_\text{crit} = 8.6\times10^{-27}$~kg~m$^{-3}$.
The total enclosed mass within the virial radius, $r_{200}$, is given by,
\begin{equation}
M_{200} = \frac{800 \pi \rho_\text{crit}}{3} \bigg( \frac{9 \beta' k_\text{b} \tau_\text{x}}{800 \pi G \bar{\mu} m_{\text{p}} \rho_\text{crit}} - r_\text{c}^2 \bigg)^{3/2} ,
\label{virial mass}
\end{equation}
where $G$ is the gravitational constant. The baryon fraction within the virial radius is found through the following ratio:
\begin{equation}
f_\text{gas,200} = \frac{4\pi \rho_\text{c}}{M_{200}} \int_{0}^{r_{200}} \big( 1 + [r/r_\text{c}]^2 \big)^{-3\beta'/2} r^2 \D r ,
\end{equation}
where the virial radius upper bound is constrained from the virial mass (\cref{virial mass} using $M_{200} = 800 \pi \rho_\text{crit} r_{200}^3/3$). The reference hydrodynamic simulation is, in this manner, found to have a virial mass of $1.34\times 10^{14}$~M$_\odot$ and a baryon fraction of 21.6\%.

We consider a further nine mock cluster environments with virial masses in the range $M_{200} \in (0.225,3.87)\times 10^{14}$~M$_\odot$ and baryon fractions $f_\text{gas,200} \in (0.098,0.153)$. The core density, core radius, slope of the density profile (i.e., $\beta'$), and isothermal temperature are modified (a few at a time) relative to the reference simulation to yield cluster environments with a virial mass and baryon fraction in the desired range. The properties of the ambient density and temperature profiles for ten clusters considered in this work are listed in \cref{tab:clusters}; we refer to these in text as Clusters A through K.

{The representative cluster environments considered here neglect cold gas and large-scale bulk motions within the intracluster medium; i.e., `cluster weather'. While relativistic jets are particularly sensitive to interactions with dense gas clouds \citep[e.g.,][]{Ehlert+2023,Seymour+2025}, the subsequent expansion of the lobe--shocked shell system is governed primarily by the pressure of the shocked plasma for supersonic expansion, rather than by interactions with individual dense gas clouds. Cluster weather may, of course, distort the shocked shell or redirect buoyant bubbles, consequently redistributing the resulting AGN heating \citep[e.g.,][]{Morsony+2010,Bourne+2021}. The results presented below should therefore be interpreted as the heating expected for idealised cluster atmospheres.}

\subsubsection{Gravity and internal energy}
\label{sec:Gravity and internal energy}

\begin{figure*}
\includegraphics[width=\textwidth,trim={95 5 112 45},clip]{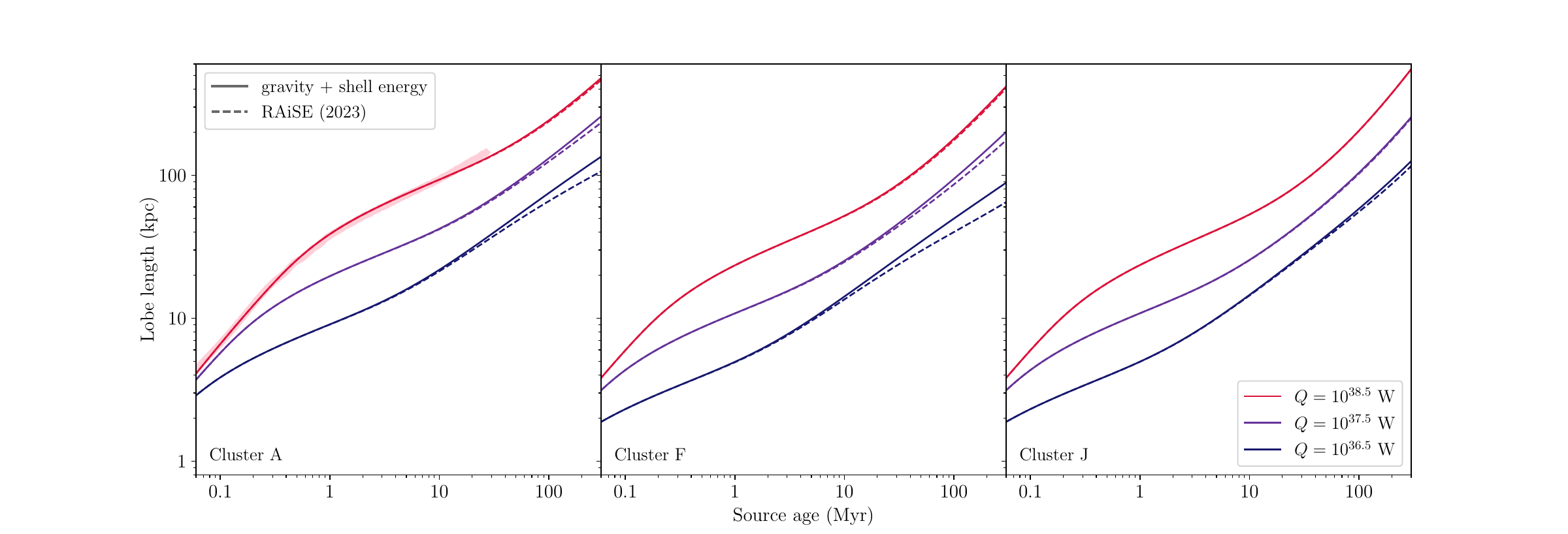}
\caption{Modelled expansion of the lobe length (single lobe) as a function of the source age using the RAiSE model either including (solid lines) or excluding (dashed lines) the gravitational and ambient internal energy corrections. The evolutionary tracks are shown for three jet powers and three cluster environments: Clusters A, F and J. The evolution of the radio source in Cluster A with a $10^{38.5}$~W jet power assumes the same intrinsic parameters as the reference hydrodynamic simulation (shown in pink, left panel only).}
\label{fig:dynamics}
\end{figure*}

The lobe length evolution of RAiSE is examined for a range of jet powers ($Q = 10^{36.5}$, $10^{37.5}$ and $10^{38.5}$~W, one-sided power) in three of the synthetic clusters (Clusters A, F and J); shown in \cref{fig:dynamics}. The predicted evolution for a $10^{38.5}$~W jet power in Cluster A agrees with that of the hydrodynamic simulation, which assumes these same jet and environment parameters{, including a Lorentz factor of $\gamma_\text{j} = 5$ along the jet spine \citep[cf.][their Equation 9b]{Turner+2023a}}. The hydrodynamic simulation, of course, considers both the change in gravitational potential energy and the internal energy of swept-up ambient gas; however, the evolutionary tracks in this region of parameter space are consistent with the simulation outputs with or without the correction term. Regardless, this result confirms that our improved RAiSE model{ remains consistent with the hydrodynamic simulation calibration in \citet{Turner+2023a}.}

The two lower-power jets in Cluster A both show a small but noticeable difference in their growth rate shortly after $t = 10$~Myr; the internal energy of the swept-up gas accelerates their expansion relative to the \citet{Turner+2023a} model. This is expected as the forward expansion speed of the lobe--shocked shell system is closer to the sound speed of the ambient medium (i.e., $v_\text{s} \approx c_\text{x}$) and, consequently, the correction term has a relatively larger effect on the acceleration (see \cref{supersonic system grav}). The same behaviour is apparent for these two jet powers in Cluster F. Clusters A and F are both massive clusters with high sound speeds of $c_\text{x} = 8.91\times 10^5$~m\,s$^{-1}$. By contrast, the cooler, lower mass Cluster J has a sound speed of $c_\text{x} = 3.98\times 10^5$~m\,s$^{-1}$ and rapidly falling density profile at higher galactocentric radii ($\beta' = 0.76$; i.e., $\beta > 2$ as $r \rightarrow \infty$); in this steep environment, the gravitational correction becomes more comparable to the ambient internal energy correction yielding $\Delta \approx 0$ (see \cref{Delta 3}). The evolutionary tracks in this cluster are, consequently, largely unaffected by the inclusion of the correction term for any of the three jet powers considered.

\subsection{Feedback energetics}

We investigate the energetics of kinetic-mode AGN feedback as a function of galactocentric radius. The heating induced during the active phase, and for several potential scenarios in the remnant phase, is considered in \cref{sec:Feedback mechanisms}. The energy coupling efficiency is modelled across active age--jet power parameter space for each of our ten mock clusters in \cref{sec:Cluster coupling efficiency}. Finally, in \cref{sec:Time-average heating rate}, we derive the time-averaged heating rate for each of our clusters across multiple outbursts and compare to the X-ray gas cooling rate.

\subsubsection{Remnant feedback mechanisms}
\label{sec:Feedback mechanisms}

The ultimate fate of the lobe--shocked gas system (i.e., upon returning to approximate hydrostatic equilibrium) is not modelled explicitly in the RAiSE framework, however, we consider the energetics of two end-member cases (see \cref{sec:shocked shell and lobe} for a complete description): 1) the underdense lobe may rise buoyantly to large galactocentric radii (and the shocked gas shell collapse inwards), or 2) the entire lobe--shocked shell system may implode upon the cessation of jet activity. In the former case, we model the effect of ablation given bubbles are typically observed within a few pressure scale heights \citep[order a few 100~kpc;][]{Churazov+2003,Zhang+2018}.


\begin{figure}
\includegraphics[width=\columnwidth,trim={12 25 48 68},clip]{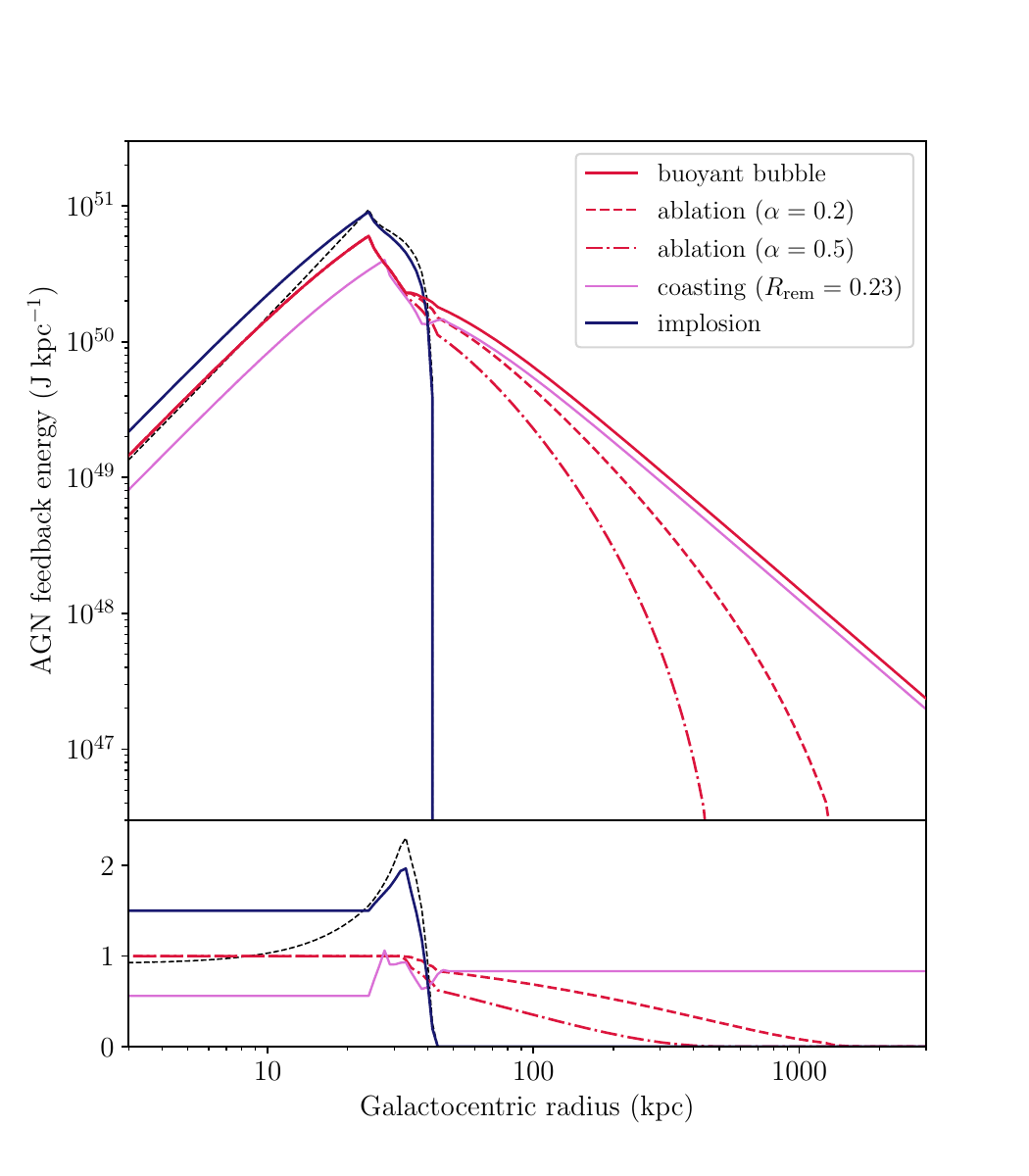}
\caption{Feedback energetics for a $Q = 10^{38}$~W jet power source with an active age of $t_\text{on} = 10$~Myr in Cluster D. \textit{Top:} The AGN heating (i.e., the increase in internal energy), integrated over polar angle and azimuth, is shown at the cessation of jet activity (black dashed), and upon returning to an approximate equilibrium assuming the lobe rises buoyantly (solid red; dashed and dot-dashed line include ablation) or implodes inwards (solid blue). The AGN heating for the perfectly adiabatic buoyant bubble is additional calculated assuming the shocked shell coasts for $t_\text{off} = 3$~Myr (solid pink). \textit{Bottom:} The fraction of AGN heating provided for each mechanism relative to the perfectly adiabatic bubble (and associated collapsing shocked gas shell).}
\label{fig:ablation}
\end{figure}

\begin{figure}
\includegraphics[width=\columnwidth,trim={24 144 64 107},clip]{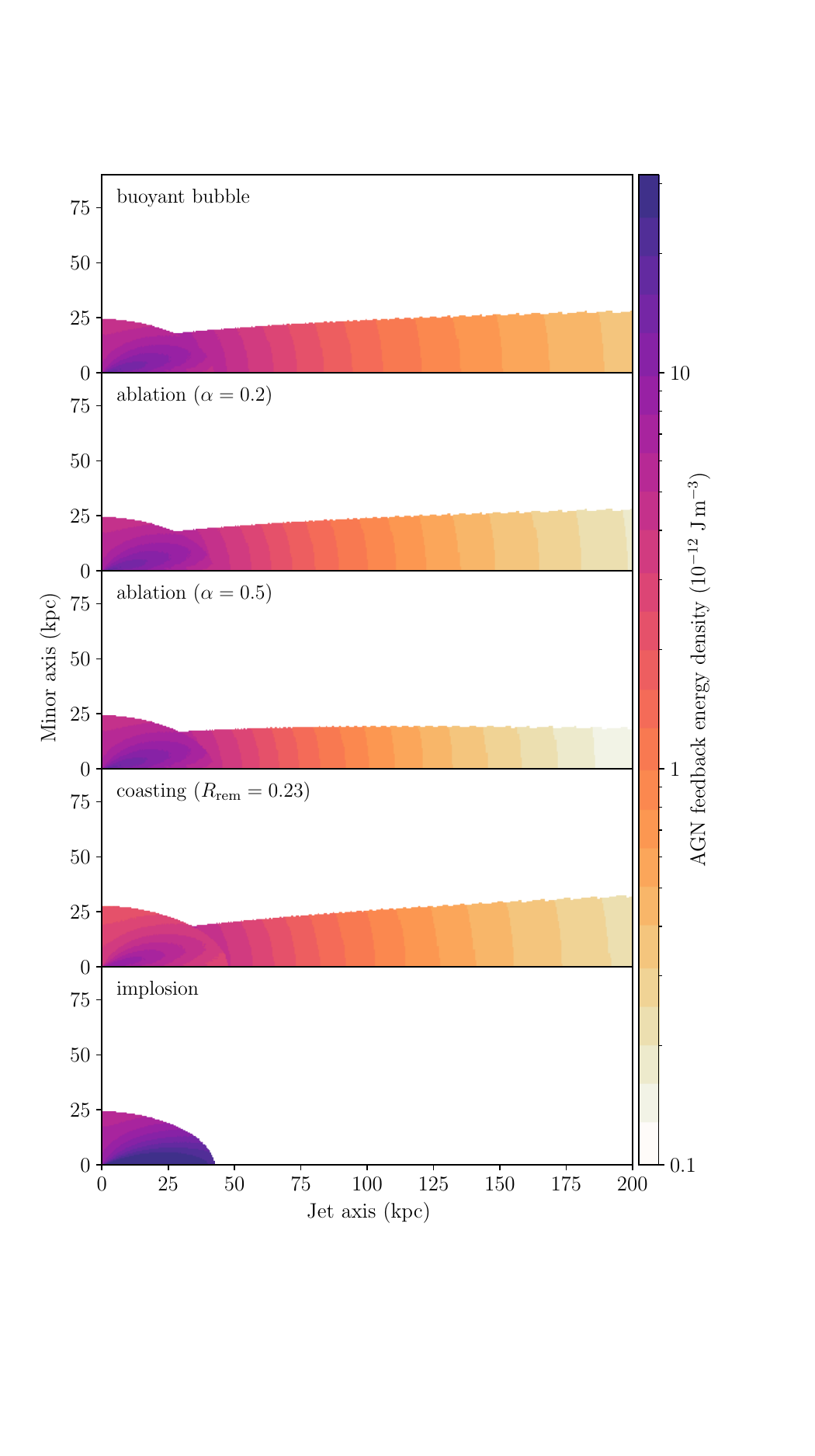}
\caption{Radial and polar-angle distribution (reprojected into the Cartesian plane) of feedback energetics for a $Q = 10^{38}$~W jet power source with an active age of $t_\text{on} = 10$~Myr in Cluster D (as for \cref{fig:ablation}). The panels show the feedback energetics for a buoyant bubble (top), gently ablated bubble ($\alpha = 0.2$; second), moderately ablated bubble ($\alpha = 0.5$; third), buoyant bubble after a 3~Myr `coasting' phase (fourth), and an implosion inwards (bottom).}
\label{fig:spatial}
\end{figure}

We model the AGN feedback energy (or heating) for each of these two processes assuming a $Q = 10^{38}$~W jet expanding in Cluster D for an active age of $t_\text{on} = 10$~Myr; shown in \cref{fig:ablation}. The AGN heating is presented as the azimuthal and polar-angle integral over the half-sphere of the cluster about the jet axis (i.e., units of Joules per kpc in the radial direction). The increase in internal energy at the end of the active phase is shown for comparison. The heating for a remnant implosion extends to the maximal extent of the active source (i.e., $r = R_\text{s}(\theta)$), as expected, but is concentrated towards lower radii when compared to the internal energy distribution at the cessation of jet activity{; we note that an implosion is implausible at the jet power considered, but include the predicted heating for completeness \citep[cf.][]{Turner+2026}}. By contrast, the buoyantly rising bubble primarily heats regions of the cluster beyond that of the active source (i.e., $r > R_\text{s}(\theta)$), while the collapsing shocked shell (associated with the buoyant bubble) provides 35\% less heating than that of the remnant implosion at lower radii (\cref{fig:ablation}, lower panel). Overall, the total heating provided by this buoyant bubble in Cluster D is 65\% greater than for the remnant implosion. 

We consider the inclusion of ablation in the model for the buoyant bubble for two spatial ablation rates: $\alpha = 0.2$ and $0.5$. 
For Cluster D, with $\beta \rightarrow 3\beta ' = 1.7$ at large galactocentric radii, the maximum radius a bubble is expected to reach is $R_\text{max} \approx \sqrt{\;\!8 V(R_\text{s})/ (R_\text{s} \alpha^3)} \approx 4R_\text{s} \alpha^{-1.5}$ (cf. \cref{Rmax}). The full model prediction in \cref{fig:ablation} compares well with this approximation: $R_\text{max} \approx 1350$ and $350$~kpc for $\alpha = 0.2$ and $0.5$, respectively, assuming a polar-averaged radius for the shock shell of $R_\text{s} \approx 30$~kpc.
The AGN feedback energy provided for the previously discussed bubble in Cluster D with these two ablation rates is shown in \cref{fig:ablation}. The AGN feedback energy is lower at all galactocentric radii, as expected, since less work is done against the gravitational potential of the cluster for the bubble with decreasing volume. Compared to the perfectly adiabatic bubble, the total heating is reduced by 30\% and 45\% for spatial ablation rates of $\alpha = 0.2$ and $0.5$, respectively. The internal energy released by the bubble as it disintegrates contributes 5\% of the total feedback energy (i.e., $\alpha p(r) A(r)$ term in \cref{u tilde 1}); this cannot compensate for the lost gravitational heating in any ambient medium with $\beta \geqslant 0$ (see \cref{u tilde 2}). We assume the limiting case of no ablation (i.e., $\alpha = 0$) throughout this work though caution that the heating at high galactocentric radii may be overestimated.

The spatial distribution of the AGN feedback energy density is plotted in \cref{fig:spatial} for the same radio source as above. We have reprojected our expressions for the energy density derived in \cref{sec:feedback energetics} from $(r,\theta)$ to the Cartesian plane. The AGN heating from the buoyant bubble is concentrated along the jet axis as expected, with a tighter range of angular values when ablation is present due to the smaller lobe volumes (at any given galactocentric radius). Meanwhile, the heating from the imploding remnant lobe is more pronounced along the jet axis than for the collapsing shocked gas shell (associated with a buoyant bubble); this occurs as the volume of the energy-dense lobe is similarly concentrated along the jet axis. The polar-angle dependence is neglected in the {remainder} of this paper for readability, however, our data products in the online supplement retain the complete spatial distribution.

We investigate `coasting' of the shocked gas shell at early-times in the remnant phase by continuing the growth of the lobe--shocked shell system to ages consistent with observed remnants with this morphology (e.g., \citealt{Quici+2022}; eastern lobe is connected to core but western lobe may have separated). These authors constrain a remnant ratio of $R_\text{rem} = t_\text{off}/(t_\text{on} + t_\text{off}) = 0.23 \pm 0.02$ using synchrotron spectral ageing models applied to 8~arcsecond resolution beam-matched observations at seven radio frequencies. This remnant ratio corresponds to an off-time of $t_\text{off} = 3$~Myr for the radio source we have discussed so far in this section. The AGN feedback energy provided for a buoyant bubble (and collapsing shocked shell) after the shock-front subsides over a further 3~Myr in a `coasting' phase is shown in \cref{fig:ablation,fig:spatial}. The total feedback energy imparted within the maximal extent of the shocked shell is comparable but spread over a larger radial range (compared to no `coasting' phase) leading to a factor of two decrease in energy density. The internal energy of the lobe is largely unchanged after 3~Myr (i.e., $U_\text{lobe} \propto p^{(\Gamma_\text{c} - 1)/\Gamma_\text{c}} \propto r^{-2\beta/5}$; see \cref{u lobe}); the gas heating at large galactocentric radii is consequently unaffected for radio sources that switch-off within the cluster core.

\subsubsection{Energy coupling efficiency}
\label{sec:Cluster coupling efficiency}

\begin{figure*}
\includegraphics[width=\textwidth,trim={98 120 189 90},clip]{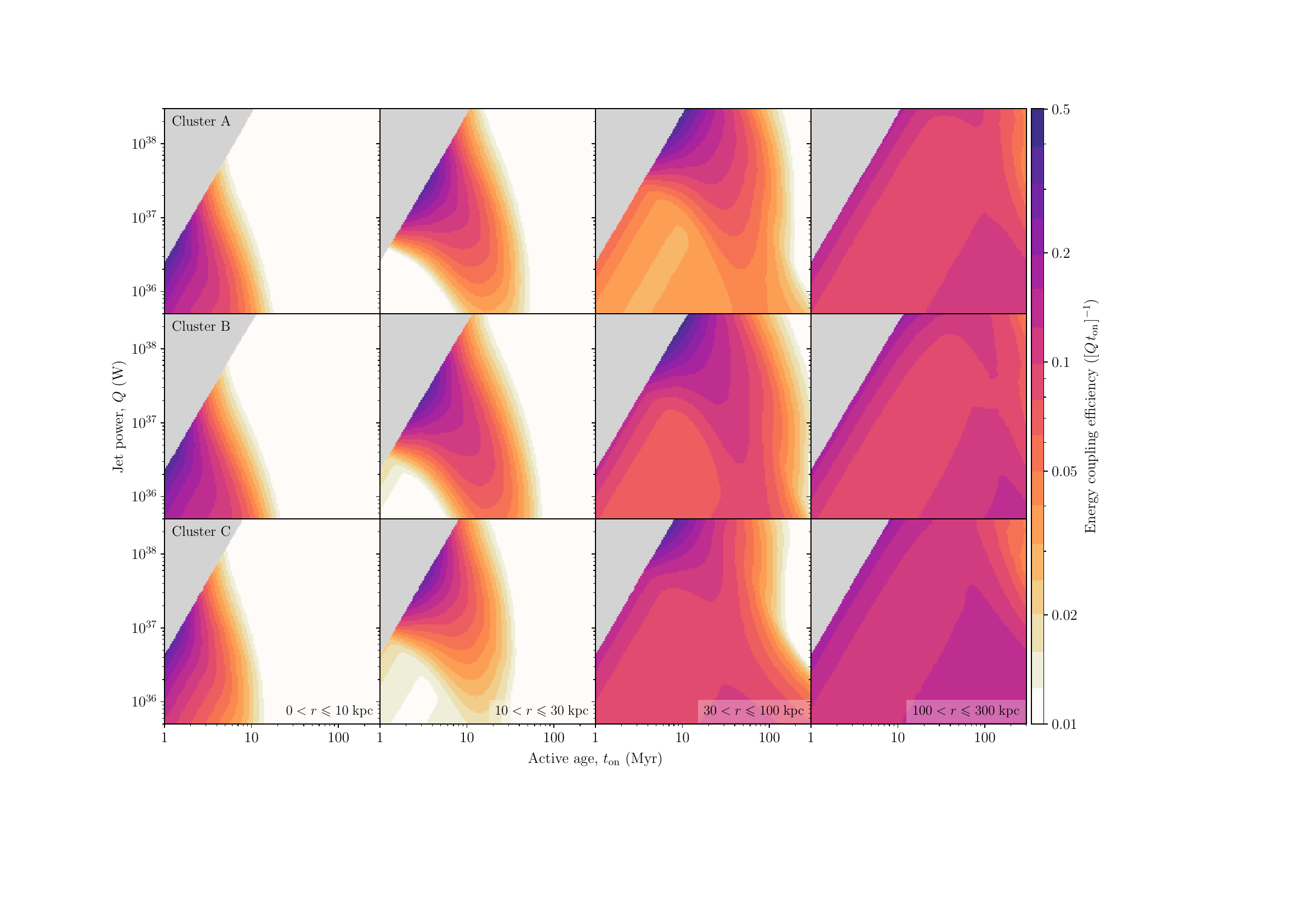}
\caption{Feedback energy coupling efficiency as a function of jet power and active age for four radial bins: $0 < r \leqslant 10$, $10 < r \leqslant 30$, $30 < r \leqslant 100$ and $100 < r \leqslant 300$~kpc (left to right). The coupling efficiency is derived for a buoyant bubble and scaled by the input energy of the outburst. The three rows show Clusters A, B and C with relatively flat density profiles at large galactocentric radii (i.e., $\beta'=0.38$).}
\label{fig:radial bands 1}
\end{figure*}

\begin{figure*}
\includegraphics[width=\textwidth,trim={98 175 189 122},clip]{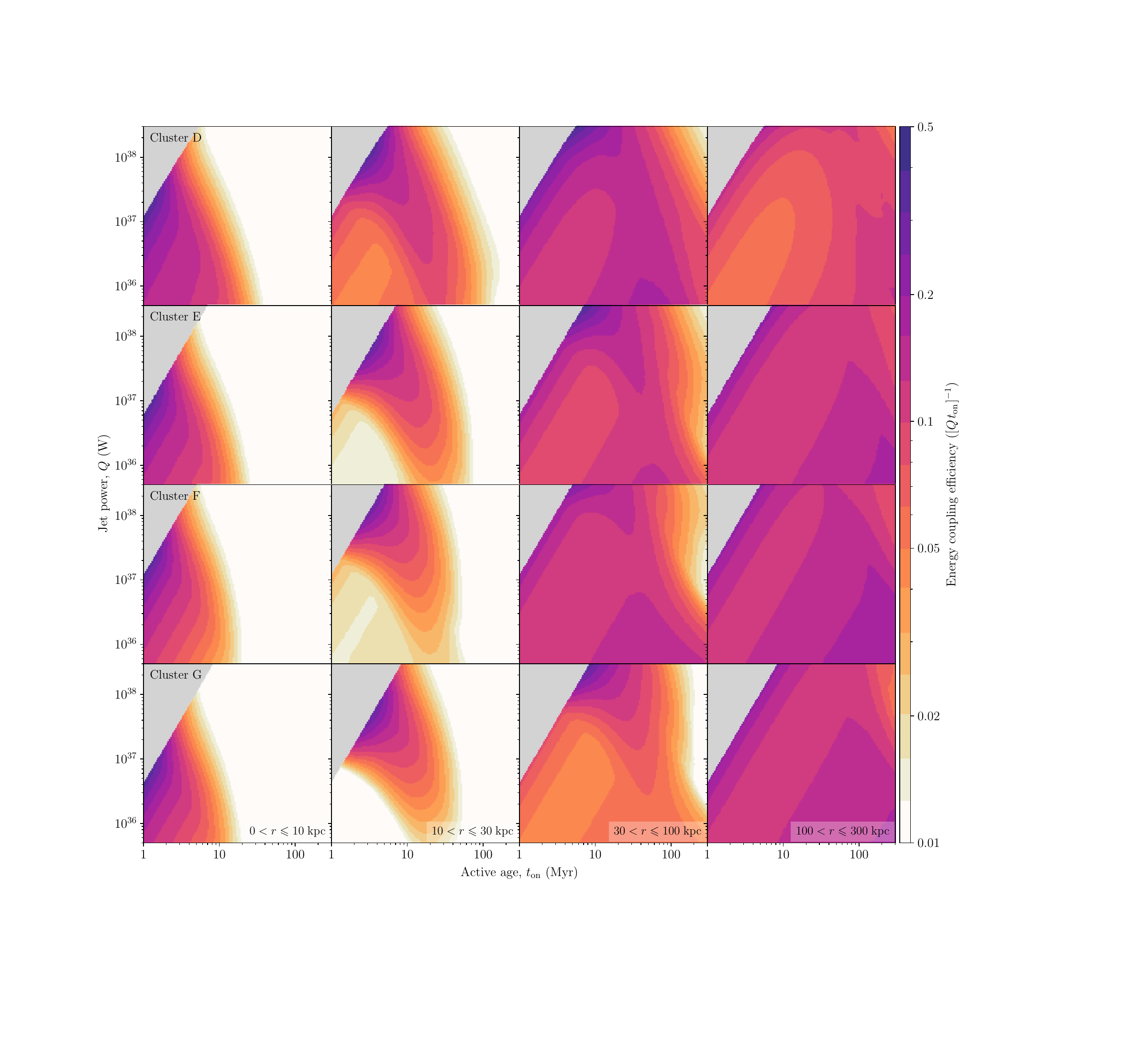}
\caption{Same as \cref{fig:radial bands 1} but for Clusters D, E, F and G with moderately steep gas density profiles at large galactocentric radii (i.e., $\beta' = 0.57$).}
\label{fig:radial bands 2}
\end{figure*}

\begin{figure*}
\includegraphics[width=\textwidth,trim={98 120 189 90},clip]{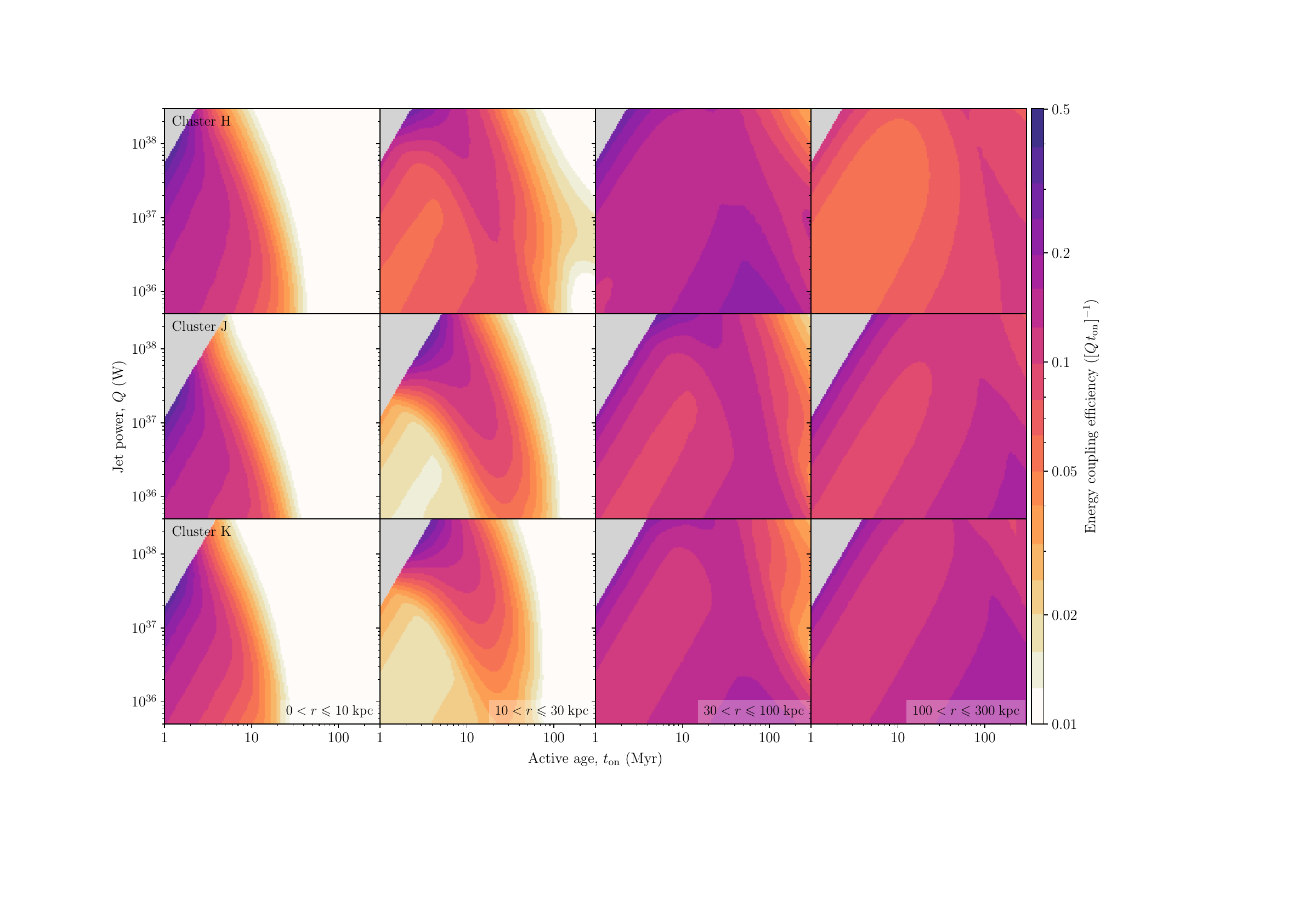}
\caption{Same as \cref{fig:radial bands 1} but for Clusters H, J and K with steep gas density profiles at large galactocentric radii (i.e., $\beta' = 0.76$).}
\label{fig:radial bands 3}
\end{figure*}

We progress our exploration of AGN feedback energetics (as in \cref{sec:Feedback mechanisms}) by covering the breadth of jet power--active age parameter space for each of our ten mock cluster environments (see \cref{sec:Cluster environments}). The high-dimensionality of this parameter space necessitates binning the galactocentric radius to effectively visualise our results; the raw data is available in full in the online supplementary material. We choose four radial bins: $0 < r \leqslant 10$, $10 < r \leqslant 30$, $30 < r \leqslant 100$ and $100 < r \leqslant 300$~kpc. The heating is integrated over polar angle and azimuth for the half-sphere of the cluster about the jet axis. The resulting radially-binned heating is scaled by the total input energy of the outburst to isolate how inherently effective different jet configurations are at heating their localised environments over time; this `energy coupling efficiency' is presented in \cref{fig:radial bands 1,fig:radial bands 2,fig:radial bands 3}.

The energy coupling efficiency is strongly dictated by the both jet power and active age. At low jet powers ($Q < 10^{37}$~W) and young active ages ($t_\text{on} < 10$~Myr), the majority of gas heating is confined to the innermost radial bin ($0<r \leqslant 10$~kpc). As the active age increases, the maximum extent of the source moves outwards with the peak coupling efficiency shifting to larger galactocentric radii (e.g., the $10 < r \leqslant 30$~kpc bin for 10-100~Myr active ages). Conversely, these older sources impart an increasing small fraction of their total energy budget within the small volumetric confines of the innermost radial bin (e.g., the volumes of the first and second radial bins differ by a factor of eight). The same behaviour is apparent at higher jet powers ($Q > 10^{37}$~W) except, of course, the lobes reach higher galactocentric radii for a given active age. Crucially, the higher jet powers delay the formation of lobes leading to some regions of the active age--jet power parameter space being occupied by partially-formed lobe structures or ballistically expanding jets. We mask these regions in \cref{fig:radial bands 1,fig:radial bands 2,fig:radial bands 3} (grey shading) as the energy calculation for a buoyant bubble is not strictly applicable.

Outside the flat cluster core, the energy coupling efficiency is heavily modulated by the density profile of the cluster. The work done per unit length is proportional to the slope of the local density profile (i.e., $\beta \approx 3 \beta'$; see \cref{u tilde 2}). The relatively flat environments of Clusters A, B and C ($\beta' = 0.38$; \cref{fig:radial bands 1}) cause the buoyant bubbles to expand slowly, and retain their structural integrity to very-high radii in the case of ablation, with only modest heating in the outermost radial bins ($r > 30$~kpc). Conversely, in Clusters H, J and K with steep density profiles (i.e., $\beta' = 0.76$; \cref{fig:radial bands 3}), the rapid decline in external pressure dramatically alters the bubble dynamics. The weaker jets achieve enhanced coupling efficiencies at large radii ($r>30$~kpc). High-powered jets are less obviously affected as their larger sizes result in the collapsing shocked gas shell depositing considerable energy in the $30 < r \leqslant 100$~kpc radial bin.

The size of the cluster core similarly affects the location of gas heating. Clusters A and G, with the largest cores ($r_\text{c} = 144$~kpc; see \cref{fig:radial bands 1,fig:radial bands 2}), have greatly reduced coupling efficiencies for weaker jets in the $30 < r \leqslant 100$~kpc radial bin. This is expected as the density profile remains locally flat (or close to flat) out to the core radius. By contrast, Clusters D and H, with the most compact cores ($r_\text{c} = 28.8$~kpc; see \cref{fig:radial bands 2,fig:radial bands 3}), have marginally enhanced gas heating in the $10 < r \leqslant 30$~kpc bin and decreased heating at the highest galactocentric radii (i.e., $r > 100$~kpc). The adiabatically expanding bubble has a significantly lower pressure at high radii in steep environments, and is consequently relatively ineffective at heating the ambient medium (i.e., $\tilde{u} \propto p$; see \cref{u tilde 2}).

\subsubsection{Time-averaged heating rate}
\label{sec:Time-average heating rate}

The heating provided by AGN feedback is sufficient to offset radiative cooling in the cores of many observed clusters \citep[e.g.,][]{Rafferty+2006,Hlavacek-Larrondo+2012,Russell+2013}. We examine the time-averaged gas heating rate provided across multiple duty cycles as a function of cluster location. The heating is compared to the gas cooing rate within the same volume of undisturbed ambient gas (i.e., \cref{cooling rate} with the gas density profile and temperature of the relevant cluster).

The histories for the active age and jet power of each outburst across multiple duty cycles are modelled using the statistics of \citet{Shabala+2020} and \citet{Quici+2025}. That is, jet power and active lifetime functions of the form
\begin{equation}
\begin{split}
&p(Q)\D Q = (Q/Q_0)^a \D Q, \\
&p(t_\text{on})\D t_\text{on} = (t_\text{on}/t_0)^b \D t_\text{on} ,
\end{split}
\end{equation}
where \citet{Quici+2025} constrain $a = -1.5$ based on a sample of 79 remnants with individual energetic estimates, \citet{Shabala+2020} fit $a = -1$ for a larger population of active and remnant sources, and both authors find $b = -1$ consistent with the expectation for self-regulated feedback \citep[cf.][]{Gaspari+2013}. We assume minimum jet powers and active ages of $10^{36}$~W and 0.3~Myr, respectively; our results are robust to variations of up to an order of magnitude in these values. The time-averaged heating rate is derived by weighting the AGN feedback energy for each active age--jet power pair (cf. \cref{fig:radial bands 1,fig:radial bands 2,fig:radial bands 3}) by the normalised active age and jet power functions, and dividing by the active age to approximate the heating rate over a single duty cycle for $\delta = 1$ (i.e., 100\% duty cycle). 

\begin{figure*}
\includegraphics[width=\textwidth,trim={95 2 112 35},clip]{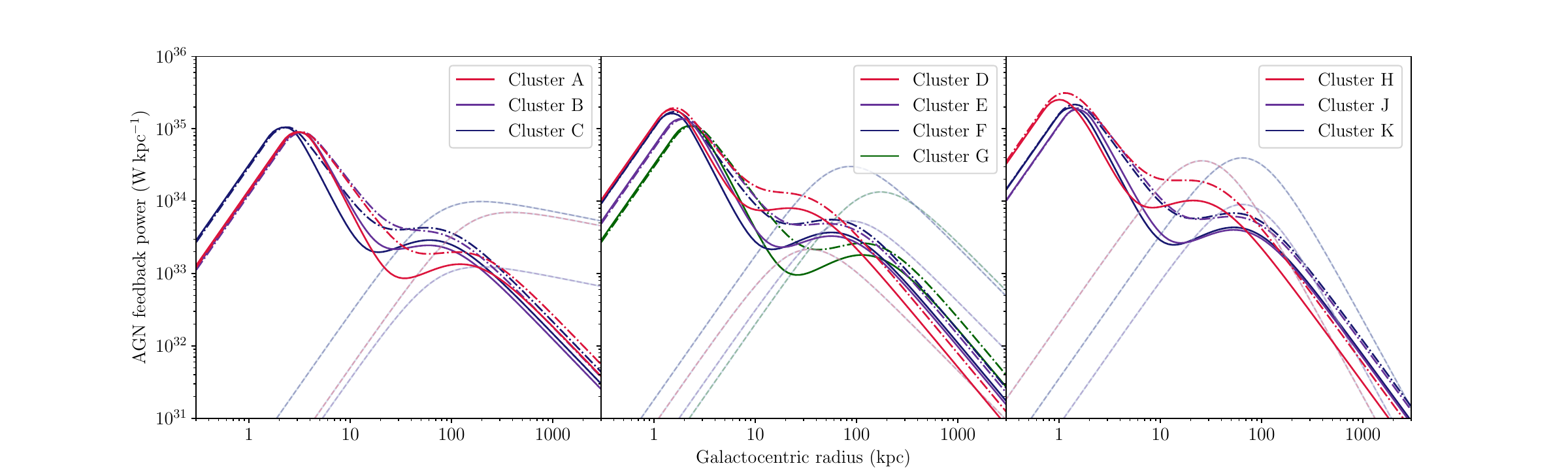}
\caption{The AGN heating rate (single lobe/half-sphere) as a function of galactocentric radius averaged across numerous jet outbursts for a 100\% duty cycle. The statistics of each outburst follow the active age and jet power functions of either \citet[][solid lines]{Quici+2025} or \citet[][dot-dashed lines]{Shabala+2020}. The clusters with flatter gas density profiles at large radii ($\beta' = 0.38$) are shown in the left panel, those with $\beta' = 0.57$ in the centre panel, and the steepest environments ($\beta' = 0.76$) in the right panel. The cooling rate for the undisturbed ambient gas in each cluster is shown by the dashed (and blue shaded) lines.}
\label{fig:time average}
\end{figure*}

The time-averaged gas heating rates are shown in \cref{fig:time average} assuming a perfectly adiabatic buoyant bubble. We considered modelling the heating rate of a remnant implosion for active age--jet power pairs below the critical line proposed by \citet{Turner+2026} but found minimal difference as this region of parameter space is associated with low jet powers (and thus heating rates) and, further, is down-weighted by the typically large active ages. The heating rate shows a peak of 1-$2\times 10^{35}$~W\;kpc$^{-1}$ at 1-3~kpc in all clusters; this is a radial heating rate, with the volumetric heating rate constant up to this radius. The radial heating rate reduces rapidly from 1-3 to 10~kpc as fewer of the ubiquitous weak and short-lived sources occupy the full spherical volume of a given radial bin. The flattening towards 100~kpc corresponds to the increasing energy coupling efficiency of the buoyant bubble as the ambient medium begins to steepen in most clusters. The flattening is delayed in Clusters A and G with the largest core radii ($r_\text{c} = 144$~kpc), and earliest in Clusters D and H with the most compact cores ($r_\text{c} = 28.8$~kpc). The feedback power is a factor of a few more effective near the core radius in steeper environments ($\beta' = 0.76$; right panel of \cref{fig:time average}) than in flatter clusters ($\beta' = 0.38$; left panel). The heating rate beyond the cluster core reduces in proportion to the ambient pressure (i.e., $\tilde{u} \propto p$; see \cref{u tilde 2}), falling more rapidly in the steeper cluster environments.

\begin{table}
\begin{center}
\newcolumntype{L}{>{\raggedright\arraybackslash}m{100pt}}
\caption[]{Time- and spatially-averaged volumetric heating and cooling rates within the innermost 100~kpc of each cluster (second and third columns). The heating rate is presented assuming the statistics of \citet{Quici+2025} and a 100\% duty cycle. The minimum duty cycle, $\delta$, that provides sufficient AGN heating to offset radiative cooling is listed in the fourth column; $\dagger$ indicates that AGN heating cannot offset cooling for any duty cycle.}
\label{tab:heating/cooling}
\renewcommand{\arraystretch}{1.1}
\setlength{\tabcolsep}{6pt}
\begin{tabular}{cccc}
\hline\hline
Cluster&Heating rate&Cooling Rate&Duty Cycle, $\delta$ \\
&($\times 10^{-30}$~W\;m$^{-3}$)&($\times 10^{-30}$~W\;m$^{-3}$)&\\
\hline
A & 7.90  & 1.96  & 0.25 \\
B & 9.81  & 0.843  & 0.086 \\
C & 9.17  & 6.74  & 0.74 \\
D & 14.38  & 2.37  & 0.16 \\
E & 10.51  & 5.37  & 0.51 \\
F & 10.54  & 30.37  & 2.9$^\dagger$ \\
G & 7.80  & 6.92  & 0.89 \\
H & 15.82  & 30.02  & 1.9$^\dagger$ \\
J & 11.88  & 9.85  & 0.83 \\
K & 11.78  & 43.27  & 3.7$^\dagger$ \\
\hline
\end{tabular}
\end{center}
\end{table}

\begin{figure}
\centering
\includegraphics[width=\columnwidth,trim={0 2 37 35},clip]{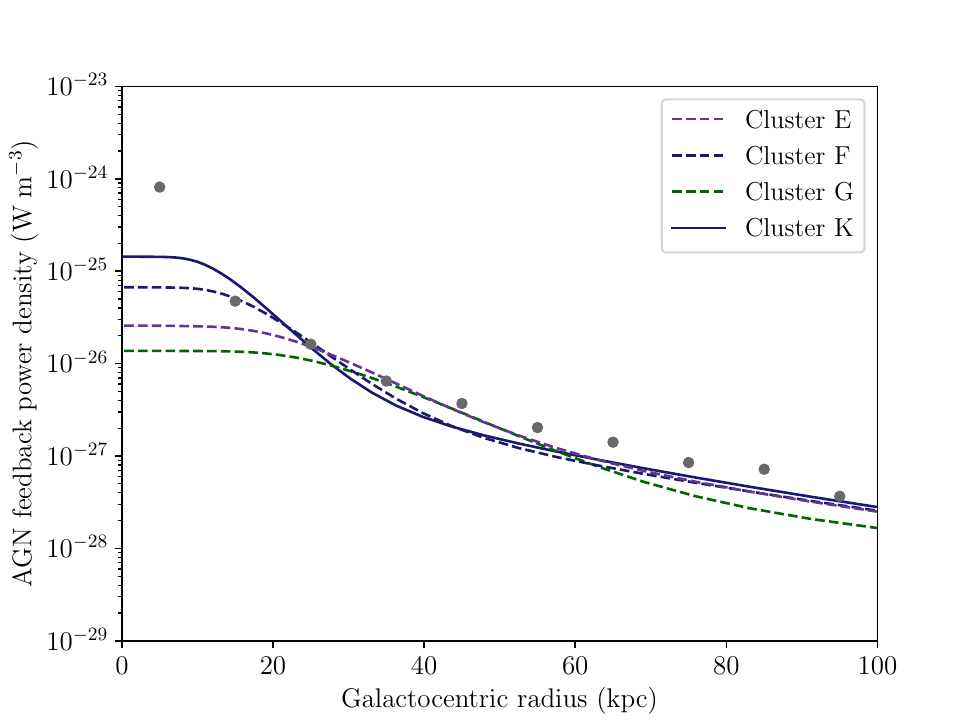}
\caption{{Comparison of the predicted AGN volumetric heating rate (single lobe/half-sphere) for a jet power of $Q=10^{38.2}$~W with the hydrodynamic simulations of \citet{Yang+2016}. The grey points are the polar-averaged `shock heating' and `transport+adiabatic' heating rates from \citet{Yang+2016}, while the coloured curves show our predicted heating rates for the four cluster atmospheres in this work that most closely resemble their simulated Perseus-like cluster.}}
\label{fig:time average comp}
\end{figure}

{The self-regulated AGN feedback simulation of \citet{Yang+2016} provides a useful validation of our heating rates. The time-averaged total jet power in their simulation is $3\times10^{45}\,\mathrm{erg\,s^{-1}}$, corresponding to a single-sided jet power of $Q=10^{38.2}$~W, which we represent using a log-space Gaussian jet power distribution of width $\sigma = 0.5$~dex. Their Perseus-like cluster atmosphere is most closely represented by our more massive cluster environments: Clusters E, F, G and K. We compare our predicted heating rates with the polar-averaged `shock heating' and `transport+adiabatic' heating rates across the jet cones and ambient medium, extracted from Figure 11 of \citet{Yang+2016}. Our radial heating rates are in good agreement with the simulated heating rates, reproducing both the magnitude and radial dependence of the deposited energy; see \cref{fig:time average comp}. The hydrodynamic simulation predicts a higher heating rate within the innermost radial bin as our active lifetime function has a minimum age of 0.3~Myr. We proceed to analyse our full model predictions using the previously adopted observationally informed jet power functions.}

{In \cref{fig:time average}, we see that the} heating rate exceeds the cooling rate by three to four orders of magnitude at small galactocentric radii (i.e., $r < 1$-3~kpc) in all clusters, reaching parity in most clusters between 10 and 100~kpc. The (time- and) spatially-averaged volumetric heating and cooling rates within the inner 100~kpc of each cluster are listed in \cref{tab:heating/cooling}. The heating provided by AGN feedback is sufficient to offset cooling in this cluster core region for seven of the ten clusters with duty cycles $0.08 < \delta \leqslant 1$, consistent with the observed radio-loud fraction \citep[e.g.,][]{Best+2005,Sabater+2019}; we caution that a `coasting' phase (see \cref{sec:Feedback mechanisms}) will lower the heating rate in the cluster core and, consequently, require a higher duty cycle for heating--cooling balance. However, even for a 100\% duty cycle (and no `coasting'), the time-averaged AGN heating in Clusters F, H and K cannot offset cooling in the cluster core. These three clusters have the highest gas cooling rates, primarily driven by their very high core densities ($\rho_\text{c} \geqslant 14.5\times 10^{-24}$~kg\;m$^{-3}$; cf. \cref{cooling rate}). Compared to our predictions, core cooling will be somewhat reduced, as each AGN outburst pushes the most rapidly cooling ambient gas out to larger galactocentric radii. However, since the peak (in energy) of the radial cooling function occurs at approximately 100~kpc (\cref{fig:time average}), where the lobe subtends only a small solid angle, this effect is likely to be relatively minor.


At large galactocentric radii ($r > 100$~kpc), the cooling rate in seven of the ten clusters exceeds the energy input from AGN heating, even under a 100\% duty cycle (\cref{fig:time average}). Clusters F, H and K, those with the highest core cooling rates, also exhibit elevated cooling at large radii, driven by their high central gas densities. Meanwhile, Clusters A, B, C and G either have relatively a flat density profile beyond the core (i.e., $\beta' = 0.38$) or the largest core radius of $r_\text{c} = 144$~kpc. The high gas densities lead to rapid cooling out to high radii, while the flatter environments prevent effective heating from buoyant bubbles. The AGN heating (for $\delta = 1$) at least balances the radiative cooling in the remaining three clusters, however, at such radii the heating is highly directional (small range of polar angles) likely leading to thermally-driven inflows and outflows. Non-AGN heating at these larger radii, such as gravitationally-driven mergers and accretion, will of course further complicate this picture \citep[e.g.,][]{Sullivan+2024}.


\section{Conclusion}
\label{sec:conclusion}

We have presented an analytic framework to derive the radial and polar-angle distribution of feedback energy from lobed AGNs in a general cluster ambient medium. The \textit{Radio AGN in Semi-analytic Environments} \citep[RAiSE;][]{Turner+2023a} dynamical model is modified to consider the work done against the gravitational potential of the cluster, the internal energy of swept-up ambient gas consumed by the bow shock, and bremsstrahlung cooling of dense gas in the resulting shocked shell (\cref{sec:lobe dynamical model}). These factors do not significantly affect the dynamics of active radio lobes except for old sources ($>30$~Myr) in flatter cluster environments (i.e., $\beta' \leqslant 0.57$; see \cref{fig:dynamics}); the internal energy of the swept-up gas is the dominant mechanism, leading to a marginally faster growth rate. The predicted lobe length evolution is consistent with a hydrodynamic simulation with the same jet parameters and ambient medium (\cref{sec:lobe--shocked shell dynamics}).

We analyse the state of the ambient medium of the large-scale cluster environment well-after the cessation of jet activity for two plausible scenarios (\cref{sec:feedback energetics}). The underdense lobe may either push through the shocked gas shell rising buoyantly to large radii, or collapse inwards as described by \citet{Turner+2026}. We consider an adiabatic bubble subject to ablation at the rear of the lobe. The perfectly adiabatic bubble (i.e., $\alpha = 0$) primarily heats regions of the cluster beyond the maximum extent of the active source (\cref{fig:ablation}; see \cref{sec:Feedback mechanisms}); increased spatial ablation rates $\alpha$ reduce the effectiveness of the bubble feedback and imposes a maximum radius of energy deposition, $R_\text{max}$ (\cref{Rmax}). The gravitational collapse of the shocked gas shell, or implosion of the lobe--shocked shell system \citep[cf.][]{Turner+2026}, leads to a significant increase in internal energy within the confines of the previous active source (\cref{fig:ablation}).

The radial and polar dependence of the AGN feedback is investigated as a function of active age and jet power for ten representative cluster environments (\cref{tab:clusters}; see \cref{sec:Cluster environments}). We derive an energy coupling efficiency -- radially-binned heating scaled by the total input energy of the outburst -- to isolate how inherently effective different jet configurations are at heating their surroundings (\cref{sec:Cluster coupling efficiency}). We compare the radiative cooling rates of our clusters to a time-averaged heating rate across multiple duty cycles based on probability density functions for the active age and jet power of each outburst \citep[][see \cref{sec:Time-average heating rate}]{Shabala+2020,Quici+2025}{; these predictions are consistent with hydrodynamic simulations of self-regulated jet feedback (\citealt{Yang+2016}; see \cref{fig:time average comp})}.
The key findings from our analysis are as follows:
\setitemize{topsep=3pt,parsep=0pt,itemsep=0pt,leftmargin=0pt,itemindent=1.5\parindent}
\begin{itemize}    
    \item Feedback from buoyant bubbles is ineffective in locally flat regions of the ambient environment (i.e., $\beta \approx 0$), or at large radii in steeply falling density profiles (i.e., $p \approx 0$; see \cref{u tilde 2}).
    \item The collapsing shocked gas shell, or imploding remnant, is responsible for heating in the flat cluster core; this is confined to the inner 10~kpc of the cluster for weak ($Q < 10^{37}$~W) and short-lived ($t_\text{on} < 10$~Myr) radio AGNs (\cref{sec:Cluster coupling efficiency}).
    \item Radio AGNs with $t_\text{on} \geqslant 100$~Myr active ages deposit less 1\% of their input energy in the inner 30~kpc of the cluster; those with $t_\text{on} \geqslant 20$~Myr deposit less than 1\% within 10~kpc (\cref{sec:Cluster coupling efficiency}).
    \item The time-averaged AGN heating rate -- due to the shocked gas shell -- falls rapidly with radius from the cluster centre as few outbursts reach larger sizes based on the statistics of \citet{Shabala+2020} and \citet{Quici+2025}; see \cref{fig:time average}. 
    \item The radial location of most effective bubble heating depends strongly on the core radius, $r_\text{c}$, and slope of the density profile beyond the core, $\beta'$; heating is most effective near the core radius for steeper environments (\cref{fig:time average}; see \cref{sec:Time-average heating rate}).
    \item Cluster cores (inner 100~kpc) with all but the highest core densities are in long-term heating--cooling balance for duty cycles $0.08 < \delta \leqslant 1$, consistent with the observed radio-loud fraction (\citealt{Best+2005,Sabater+2019}; see \cref{sec:Time-average heating rate}).
\end{itemize}

This framework to derive the radial and polar-angle distribution of gas heating rates in clusters provides a computationally effective approach to include a more complete description of AGN feedback in cosmological hydrodynamical simulations. {Within a cosmological simulation, the black hole mass, spin, and accretion rate can be used to determine the jet power, while the ambient medium can be represented by an analytic density profile (e.g., a modified $\beta$-model) describing the resolved gas distribution. The observationally informed active lifetime statistics (\cref{sec:Time-average heating rate}) provide a means to reconstruct plausible AGN outburst histories between successive simulation outputs, while conserving the total energy injected by the jet over this interval. For each outburst, the corresponding radial and polar-angle distribution of AGN heating can be evaluated and applied to the resolved gas as a physically motivated subgrid feedback prescription.}

The gas heating rates for the ten clusters considered in this work are provided in tabulated form in the online supplement; we provide this data for each active age--jet power pair ensuring some utility without running the complete computational pipeline. The analytic nature of our model avoids the need for computationally expensive runtime routines, offering a direct pathway to integrate these calculations into cosmological simulations \citep[cf.][]{Raouf+2017}. This underpins a scalable approach for modelling self-consistent jet--ambient medium interactions over cosmological timescales.

\section*{Acknowledgements}
 
{We thank the anonymous referee for their prompt and constructive review.} AS acknowledges the support of the Australian Government Research Training Program  Fees  Offset, the Bruce and Betty Green Post-graduate Research Scholarship, and The University Club of Western Australia Research Travel Scholarship. WRQG acknowledges support of the Forbes and Warren Honours Scholarship in Mathematics and Physics, and the University of Tasmania Honours Scholarship.

\section*{Data Availability}

The authors confirm that the data supporting the findings of this study are available in cited external data archives or presented in the article. Processed data products underlying this article are available in the GitHub repository: \href{https://github.com/rossjturner/feedback-energetics}{https://github.com/rossjturner/feedback-energetics}. The relevant code is publicly available as disclosed when referenced in the paper.



\bibliographystyle{mnras}
\bibliography{energetics} 





\bsp 
\label{lastpage}
\end{document}